\documentclass[a4paper,12pt,twoside]{article} 
\usepackage{a4wide}
\usepackage{graphicx}
\usepackage{rotating}
 
\newcommand{\beq}{\begin{equation}}
\newcommand{\eeq}{\end{equation}}
\def\gs{\mathrel{ \rlap{\raise
0.511ex \hbox{$>$}}{\lower 0.511ex \hbox{$\sim$}}}} \def\ls{\mathrel{
\rlap{\raise 0.511ex \hbox{$<$}}{\lower 0.511ex \hbox{$\sim$}}}}

\newcommand{\ba}{\begin{array}{c}}
\newcommand{\baz}{\begin{array}{cc}}
\newcommand{\bad}{\begin{array}{ccc}}
\newcommand{\bea}{\begin{equation} \begin{array}{c}}
\newcommand{\eea}{ \end{array} \end{equation}}
\newcommand{\ea}{\end{array}}

\def\gtap{\mathrel{ \rlap{\raise 0.511ex \hbox{$>$}}{\lower 0.511ex
   \hbox{$\sim$}}}} 
\def\ltap{\mathrel{ \rlap{\raise 0.511ex
   \hbox{$<$}}{\lower 0.511ex \hbox{$\sim$}}}}
   \newcommand{\deltaatm}{\mbox{$\Delta m^2_{\mathrm{A}}$}}
   \newcommand{\deltasol}{\mbox{$ \Delta m^2_{\odot}$}}

\begin{document}

{\flushright
Ref. SISSA 28/03/EP

FTUV-03-0513

IFIC/03-21


}

\begin{center}
\noindent{\Large \tt \bf 
Atmospheric Neutrino Oscillations, $\theta_{13}$ 
and Neutrino Mass Hierarchy 
}\vspace{4mm}
\renewcommand{\thefootnote}{\fnsymbol{footnote}}

\noindent{\large
J.~Bernab\'eu$^{1}$, Sergio Palomares-Ruiz$^1$ and
S.~T.~Petcov$^2$\footnote{Also at: Institute of Nuclear Research and
Nuclear Energy, Bulgarian Academy of Sciences, 1784 Sofia, Bulgaria.}
}\vspace{2mm}

\noindent{\small
$^1$ Departamento de F\'{\i}sica Te\'orica and IFIC, Universidad de Valencia-CSIC, 46100 
      Burjassot, Valencia, Spain 
\\
$^2$ Scuola Internazionale Superiore di Studi Avanzati 
and Istituto Nazionale di Fisica Nucleare, I-34014 Trieste, Italy}

\end{center}
\vspace{4mm}
\vspace{6mm}
\renewcommand{\thefootnote}{\arabic{footnote}}
\setcounter{footnote}{0}

\begin{abstract}
We derive predictions for the Nadir angle ($\theta_n$)
dependence of the ratio $N_{\mu}/N_e$
of the rates of the $\mu-$like and $e-$like
multi-GeV events measured in water-\v{C}erenkov detectors
in the case of 3-neutrino oscillations
of the atmospheric $\nu_e$ ($\bar{\nu}_e$)
and $\nu_\mu$ ($\bar{\nu}_\mu$), driven
by one neutrino mass squared difference,
$|\Delta m^2_{31}| \sim
(2.5 - 3.0)\times 10^{-3}~{\rm eV^2} 
\gg \Delta m^2_{21}$.
This ratio is particularly sensitive
to the Earth matter effects in the 
atmospheric neutrino oscillations,
and thus to the values
of $\sin^2\theta_{13}$ and
$\sin^2\theta_{23}$, $\theta_{13}$ and
$\theta_{23}$ being the neutrino mixing angle
limited by the CHOOZ and Palo Verde experiments
and that responsible for the dominant 
atmospheric $\nu_\mu \rightarrow \nu_{\tau}$ 
($\bar{\nu}_\mu \rightarrow \bar{\nu}_{\tau}$)
oscillations. It is also sensitive  
to the type of neutrino mass 
spectrum which can be with normal 
($\Delta m^2_{31} > 0$) 
or with inverted  
($\Delta m^2_{31} < 0$) hierarchy.
We show that for $\sin^2\theta_{13} \gtap 0.01$,
$\sin^2\theta_{23}\gtap 0.5$ and 
at $\cos\theta_n \gtap 0.4$,
the Earth matter effects 
modify substantially
the $\theta_n-$ dependence of 
the ratio $N_{\mu}/N_e$ and in a way 
which cannot be reproduced with   
$\sin^2\theta_{13} = 0$
and a different value of $\sin^2\theta_{23}$.
For normal hierarchy the effects can be as large as 
$\sim 25\%$ for $\cos\theta_n \sim (0.5 - 0.8)$,
can reach $\sim 35\%$ in the Earth core bin
$\cos\theta_n \sim (0.84 - 1.0)$,
and might be observable. They are typically by
$\sim 10\%$ smaller in the inverted hierarchy case. 
An observation of the Earth matter effects 
in the Nadir angle distribution of the 
ratio $N_{\mu}/N_e$
would clearly indicate that
$\sin^2\theta_{13} \gtap 0.01$ and 
$\sin^2\theta_{23} \gtap 0.50$.

\end{abstract}
\vspace{20pt}

\newpage
\section{Introduction}
\hskip 0.6truecm 
The publication of the first results
of the KamLAND experiment marks the beginning
of a new era in the studies of neutrino mixing 
and oscillations - the era of high precision
determination of the neutrino mixing parameters.
The data obtained by the solar  
neutrino experiments Homestake, Kamiokande, SAGE, 
GALLEX/GNO and Super-Kamiokande (SK) \cite{Cl98,SKsol}
provided the first strong evidences
for oscillations of flavour (electron) neutrinos.
Strong evidences for oscillations of the
atmospheric $\nu_{\mu}$ ($\bar{\nu}_{\mu}$)
neutrinos were obtained by 
the Super-Kamiokande (SK) 
experiment \cite{SKatm00}.
The evidences for solar 
$\nu_e$ oscillations 
into active neutrinos $\nu_{\mu,\tau}$,
were significantly reinforced during 
the last two years
i) by the combined first data of the 
SNO experiment \cite{SNO1},
and the SK data \cite{SKsol}, 
ii) by the more recent SNO neutral 
current data \cite{SNO2},
and iii) by the first results 
of the KamLAND \cite{KamLAND} experiment.
The KamLAND data practically 
establishes \cite{KamLAND},
under the plausible 
assumption of CPT-invariance, the
large mixing angle (LMA)
MSW solution as unique solution
of the solar neutrino problem.
This result brings us, after more than 
30 years of research, 
initiated by the pioneer
works of B. Pontecorvo \cite{Pont4667} and the
experiment of R. Davis et al. \cite{Davis68},
very close to a complete understanding of the 
true cause of the solar neutrino problem. 

    The interpretation of the solar and
atmospheric neutrino, and of the KamLAND
data in terms of 
neutrino oscillations requires
the existence of 3-neutrino mixing
in the weak charged lepton current 
(see, e.g., \cite{BGG99}):
\begin{equation}
\nu_{l \mathrm{L}}  = \sum_{j=1}^{3} U_{l j} \, \nu_{j \mathrm{L}}~.
\label{3numix}
\end{equation}
\noindent Here $\nu_{lL}$, $l  = e,\mu,\tau$,
are the three left-handed flavor 
neutrino fields,
$\nu_{j \mathrm{L}}$ is the 
left-handed field of the 
neutrino $\nu_j$ having a mass $m_j$
and $U$ is the Pontecorvo-Maki-Nakagawa-Sakata (PMNS)
neutrino mixing matrix \cite{BPont57},
\begin{equation}
U = \left(\begin{array}{ccc}
U_{e1}& U_{e2} & U_{e3} \\
U_{\mu 1} & U_{\mu 2} & U_{\mu 3} \\
U_{\tau 1} & U_{\tau 2} & U_{\tau 3} 
\end{array} \right)
= \left(\begin{array}{ccc} 
c_{12}c_{13} & s_{12}c_{13} & s_{13}e^{-i\delta}\\
 - s_{12}c_{23} - c_{12}s_{23}s_{13}e^{i\delta} & 
c_{12}c_{23} - s_{12}s_{23}s_{13}e^{i\delta} & s_{23}c_{13}\\
s_{12}s_{23} - c_{12}c_{23}s_{13}e^{i\delta} 
& -c_{12}s_{23} - s_{12}c_{23}s_{13}e^{i\delta}
& c_{23}c_{13}\\ 
\end{array} \right)
\label{Umix}
\end{equation}

\noindent where we have used a standard parametrization 
of $U$ with the usual notations, $s_{ij} \equiv \sin \theta_{ij}$,
$c_{ij} \equiv \cos \theta_{ij}$, 
and $\delta$ is the Dirac CP-violation phase
\footnote{We have not 
written explicitly the two possible Majorana CP-violation phases 
which do not enter into the expressions for the oscillation 
probabilities of interest 
\cite{BHP80} (see also \cite{BiPet87}). 
We assume throughout this study $0 \leq \theta_{12}, \theta_{23}, 
\theta_{13} < \pi/2$. 
}.
If we identify $\Delta m^2_{21}$ and $\Delta m^2_{31}$  
with the neutrino mass squared differences
which drive the solar and atmospheric 
neutrino oscillations, $\deltasol = \Delta m^2_{21} > 0$,
$\deltaatm = \Delta m^2_{31}$,
the data suggest that 
$|\Delta m^2_{31}| \gg \Delta m^2_{21}$.
In this case $\theta_{12}$ and $\theta_{23}$, 
represent the neutrino mixing 
angles responsible for the
solar and atmospheric 
neutrino oscillations,
$\theta_{12} = \theta_{\odot}$, 
$\theta_{23} = \theta_{\rm A}$, 
while $\theta_{13}$ is the angle 
limited by the data from
the CHOOZ and Palo Verde experiments 
\cite{CHOOZ,PaloV}. 

 Combined $\nu_e \rightarrow \nu_{\mu (\tau)}$
and $\bar{\nu}_e \rightarrow \bar{\nu}_{\mu (\tau)}$
oscillation analyses of the solar neutrino 
and KamLAND \cite{KamLAND} data,
performed under the assumption
of CPT-invariance which we will suppose to hold
throughout this study, 
show \cite{fogli,others}
that the data favor the 
LMA MSW solution with 
$\Delta m^2_{\odot} \cong 
(6.9 - 7.3)\times 10^{-5}~{\rm eV^2}$ and 
$\tan^2\theta_{\odot} \cong (0.42 - 0.46)$.
A second, statistically somewhat less favored
LMA solution (LMA II) exists 
at \cite{fogli,others} $\Delta m^2_{\odot} 
\cong 1.5\times 10^{-4}~{\rm eV^2}$.
The atmospheric neutrino data, 
as is well known,
is best described \cite{SKatm00} in terms of
dominant $\nu_{\mu} \rightarrow \nu_{\tau}$
($\bar{\nu}_{\mu} \rightarrow \bar{\nu}_{\tau}$)
oscillations with 
$|\deltaatm| 
\cong 2.5\times 10^{-3}~{\rm eV^2}$ and
$\sin^22\theta_{\rm A} \cong 1.0$.
The 90\% C.L. allowed
intervals of values of the two-neutrino 
oscillation parameters found in \cite{SKatm00}
read $|\deltaatm| 
\cong (1.9 - 4.0) \times 10^{-3}~{\rm eV^2}$ and
$\sin^22\theta_{\rm A} \cong (0.89 - 1.0)$.
According to the more recent 
combined analysis of the data from 
the SK and K2K experiments \cite{fogliold},
one has $|\deltaatm|
\cong (2.7 \pm 0.4)\times 10^{-3}~{\rm eV^2}$. 
We will often use in our analysis  
as illustrative the values 
$|\deltaatm| = 3.0\times 10^{-3}~{\rm eV^2}$ 
and $\sin^22\theta_{\rm A} = 0.92;~ 1.0$.
Let us note that the 
atmospheric neutrino and K2K data
do not allow one to determine the signs
of $\deltaatm$, and of 
$\cos2\theta_{\rm A}$ when
$\sin^22\theta_{\rm A} \neq 1.0$.
This implies that 
in the case of 3-neutrino mixing 
one can have $\Delta m^2_{31} > 0$
or $\Delta m^2_{31} < 0$. The two 
possibilities correspond to two different
types of neutrino mass spectrum:
with normal hierarchy (NH),
$m_1 < m_2 < m_3$, and 
with inverted hierarchy (IH),
$m_3 < m_1 < m_2$.  
The fact that the sign of 
$\cos2\theta_{\rm A}$ 
is not determined when
$\sin^22\theta_{\rm A} \neq 1.0$ implies
that when, e.g., 
$\sin^22\theta_{\rm A} \equiv \sin^22\theta_{23} = 0.92$,
two values of  $\sin^2\theta_{23}$ are possible,
$\sin^2\theta_{23} \cong 0.64~{\rm or}~ 0.36$.

  In what regards the mixing angle $\theta_{13}$,
a 3-$\nu$ oscillation analysis of the CHOOZ data 
\cite{BNPChooz} led to the conclusion that 
for $\deltasol  \ltap 10^{-4}~{\rm eV^2}$,
the limits on $\sin^2\theta_{13}$ 
practically coincide with
those derived in the 2-$\nu$ 
oscillation analysis in \cite{CHOOZ}.
A combined 3-$\nu$ oscillation analysis of the 
solar neutrino, CHOOZ and the 
KamLAND data, performed
under the assumption of 
$\deltasol \ll |\deltaatm|$
(see, e.g., \cite{BGG99,ADE80}),
showed \cite{fogli}
that $\sin^2\theta_{13} < 0.05$ at 99.73\% C.L.
The authors of \cite{fogli} found 
the best-fit value of $\sin^2\theta_{13}$
to lie in the interval 
$\sin^2\theta_{13} \cong (0.00 - 0.01)$.

   Getting more precise 
information about the value 
of the mixing angle $\theta_{13}$,
determining the sign of $\deltaatm$, or
the type of the neutrino mass spectrum 
(with normal or inverted hierarchy), 
and measuring the value of 
$\sin^2\theta_{23}$ with a 
higher precision
is of fundamental importance
for the progress in the 
studies of neutrino mixing. 

   The mixing angle $\theta_{13}$, or
the absolute value
of the element $U_{e3}$ of the PMNS matrix,
$|U_{e3}| =\sin\theta_{13}$, 
plays a very important  
role in the phenomenology of 
the 3-neutrino oscillations.
It drives the sub-dominant 
 $\nu_{\mu} \leftrightarrow \nu_e$
($\bar{\nu}_{\mu} \leftrightarrow \bar{\nu}_e$)
oscillations of the atmospheric 
$\nu_{\mu}$ ($\bar{\nu}_{\mu}$) and
$\nu_e$ ($\bar{\nu}_e$) \cite{SP3198,102,104}. 
The value of $\theta_{13}$   
controls also the 
$\nu_{\mu} \rightarrow \nu_e$,
$\bar{\nu}_{\mu} \rightarrow \bar{\nu}_e$,
$\nu_{e} \rightarrow \nu_{\mu}$ and
$\bar{\nu}_{e} \rightarrow \bar{\nu}_{\mu}$
transitions in the long baseline neutrino 
oscillation experiments (MINOS, CNGS),
and in the widely discussed
very long baseline neutrino oscillation 
experiments at neutrino factories (see, e.g.,
\cite{LBL,AMMS99,mantle,mantleproc12}). The magnitude of the 
T-violating and CP-violating probabilities
in neutrino oscillations is directly proportional
to $\sin\theta_{13}$ 
(see, e.g., \cite{3nuKP88,CPother,Dick99}).
Thus, in the sub-dominant channels of interest to 
T- and CP-violation 
studies, the corresponding asymmetries
become proportional to $\Delta m^{2}_{\odot}/\sin{\theta_{13}}$ 
\cite{Dick99,CPT}. 
The value of $\sin{\theta_{13}}$ is thus of prime importance.

 If the neutrinos with definite mass 
are Majorana particles (see, e.g., \cite{BiPet87}),
the predicted value of the effective Majorana mass
parameter in neutrinoless double $\beta-$decay
depends strongly in the case of 
hierarchical neutrino mass
spectrum on the value of $\sin^2\theta_{13}$ 
(see, e.g., \cite{BPP1}).   

   The sign of $\deltaatm$ determines, for instance,
which of the transitions (e.g., of atmospheric neutrinos)
$\nu_{\mu} \rightarrow \nu_e$ and
$\nu_{e} \rightarrow \nu_{\mu}$, or
$\bar{\nu}_{\mu} \rightarrow \bar{\nu}_e$ and
$\bar{\nu}_{e} \rightarrow \bar{\nu}_{\mu}$,
can be enhanced by the Earth matter effects
\cite{LW78,BPPW80,MS85}.
The predictions for the
neutrino effective Majorana mass
in neutrinoless double $\beta-$decay
depend critically on the type of the
neutrino mass spectrum 
(normal or inverted hierarchical) \cite{BPP1,BGGKP99}.
The knowledge of the value of $\theta_{13}$ 
and of the sign of $\deltaatm = \Delta m^2_{31}$
is crucial for the searches 
for the correct theory of 
neutrino masses and mixing as well.

  Somewhat better limits on $\sin^2 \theta_{13}$ than 
the existing one can be obtained in the 
MINOS experiment \cite{MINOS}. 
Various options are being currently discussed
(experiments with off-axis neutrino beams, more precise
reactor antineutrino and long base-line experiments, etc.,
see, e.g., \cite{MSpironu02}) of how to improve
by at least an order of magnitude, i.e., 
to values of $\sim 0.005$ or smaller, 
the sensitivity to $\sin^2\theta_{13}$. 
The sign of $\deltaatm$ can be determined in
very long base-line neutrino oscillation 
experiments at neutrino factories
(see, e.g., \cite{LBL,AMMS99}), and, e.g, using
combined data from long base-line
oscillation experiments at the JHF facility and
with off-axis neutrino beams \cite{HLM}.
If the neutrinos with definite mass are
Majorana particles, it can be determined
by measuring the effective neutrino Majorana
mass in neutrinoless double $\beta-$decay experiments
\cite{BPP1,BGGKP99}.
Under certain rather special
conditions it might be determined also
in experiments with reactor 
$\bar{\nu}_e$ \cite{SPMPiai01}. 

   In the present article we study possibilities
to obtain information on the value 
of $\sin^2\theta_{13}$ and on the sign of
$\deltaatm = \Delta m^2_{31}$
using the atmospheric 
neutrino data, accumulated by the SK experiment,
and more generally, that can be provided by 
future water-\v{C}erenkov detectors, like UNO and Hyper-Kamiokande.
We consider 3-neutrino oscillations 
of the atmospheric $\nu_{\mu}$, $\bar{\nu}_{\mu}$,
$\nu_e$ and $\bar{\nu}_e$ under the condition
$\deltasol = \Delta m^2_{21} \ll 
|\deltaatm| = |\Delta m^2_{31}|$,
which is suggested to hold 
by the current solar and atmospheric 
neutrino data. Under the indicated condition,
the expressions for the 
probabilities of 
$\nu_{\mu} \rightarrow \nu_e$ 
($\nu_{e} \rightarrow \nu_{\mu}$) and
$\bar{\nu}_{\mu} \rightarrow \bar{\nu}_e$
($\bar{\nu}_{e} \rightarrow \bar{\nu}_{\mu}$)
transitions contain $\sin^2\theta_{23}$ as a factor,
which determines their maximal values. 
Depending on the sign of 
$\Delta m^2_{31}$,
the Earth matter effects can resonantly enhance either
the  $\nu_{\mu} \rightarrow \nu_e$ and
$\nu_{e} \rightarrow \nu_{\mu}$,
or the $\bar{\nu}_{\mu} \rightarrow \bar{\nu}_e$
and $\bar{\nu}_{\mu} \rightarrow \bar{\nu}_{\mu}$
transitions if $\sin^2\theta_{13}\neq 0$.
The effects of the enhancement 
can be substantial for $\sin^2\theta_{13}\gtap 0.01$.
They are larger in
the multi-GeV $e$-like and $\mu-$like 
samples of events and for atmospheric 
neutrinos with relatively large path length
in the Earth, crossing deeply the mantle \cite{mantle,AMMS99}
or the mantle and the core \cite{SP3198,104,106,107,core}, i.e., for
$\cos\theta_n \gtap 0.4$, where $\theta_n$ 
is the Nadir angle characterizing 
the neutrino trajectory in the Earth.
 
   The $\nu_{\mu} \rightarrow \nu_e$ 
($\bar{\nu}_{\mu} \rightarrow \bar{\nu}_e$)
and $\nu_{e} \rightarrow \nu_{\mu}$
($\bar{\nu}_{e} \rightarrow \bar{\nu}_{\mu}$)
transitions in the Earth lead to the reduction
of the rate of the multi-GeV $\mu-$like
events and to the increase of the rate of the
multi-Gev $e-$like events in the Super-Kamiokande
(or any other water-\v{C}erenkov) detector
with respect to the case of absence
of these transitions (see, e.g., \cite{SP3198,102,104,107,core}).
Correspondingly, as observables which are sensitive 
to the Earth matter effects, and thus to the value
of $\sin^2\theta_{13}$ and the sign of 
$\Delta m^2_{31}$, as well as to 
$\sin^2\theta_{23}$, we consider the
Nadir-angle distributions of the 
ratios $N^{3\nu}_{\mu}/N^{3\nu}_{e}$ and
$N^{3\nu}_{e}/N^{0}_e$,
where $N^{3\nu}_{\mu}$ and $N^{3\nu}_{e}$ and      
are the multi-GeV $\mu-$like and  $e$-like 
numbers of events (or event rates) 
in the case of 3-$\nu$ 
oscillations of the atmospheric $\nu_e$, $\bar{\nu}_e$
and $\nu_{\mu}$, $\bar{\nu}_{\mu}$, and
$N^{0}_{e}$ is the number of $e-$like events 
in the case of absence of oscillations
($\sin^2\theta_{13} = 0$).
The ratio of the energy and Nadir angle integrated
$\mu-$like and $e-$like events, $N_{\mu}/N_e$,
has been measured with a relatively high precision by the
SK experiment \cite{SKatm00}. The systematic 
uncertainty, in particular,
in the Nadir angle dependence of the ratio
$N_{\mu}/N_e$ can be smaller than those on the measured
Nadir angle distributions of 
$\mu-$like and $e-$like events,
$N_{\mu}$ and $N_e$.

   We obtain predictions for the 
Nadir-angle distributions of
$N^{3\nu}_{\mu}/N^{3\nu}_{e}$ and of 
$N^{3\nu}_{e}/N^{0}_e$ 
both for neutrino mass spectra with 
normal ($\Delta m^2_{31} > 0$)
and inverted  ($\Delta m^2_{31} < 0$) hierarchy,
$(N^{3\nu}_{\mu}/N^{3\nu}_{e})_{\rm NH}$, 
$(N^{3\nu}_{\mu}/N^{3\nu}_{e})_{\rm IH}$,
$(N^{3\nu}_{e}/N^{0}_{e})_{\rm NH}$ and
$(N^{3\nu}_{\mu}/N^{0}_{e})_{\rm IH}$, 
and for $\sin^2\theta_{23} = 0.64;~0.50;~0.36$.
We compare the latter with the predicted
Nadir-angle distributions i) of the ratio
$N_{\mu}/N_{e}$
for the case the 3-neutrino
oscillations taking place in vacuum,
$(N^{3\nu}_{\mu}/N^{3\nu}_{e})_{\rm vac}$, and
ii) of the ratio $N^{2\nu}_{\mu}/N^{0}_{e}$,
where $N^{2\nu}_{\mu}$
is the predicted number of $\mu-$like
events in the case of 2-neutrino
$\nu_{\mu} \rightarrow \nu_{\tau}$ and
$\bar{\nu}_{\mu} \rightarrow \bar{\nu}_{\tau}$
oscillations of the atmospheric 
$\nu_{\mu}$ and $\bar{\nu}_{\mu}$.
Predictions for the different types of 
ratios indicated 
above of the suitably integrated
Nadir angle distributions
of the $\mu-$like and $e-$like 
multi-GeV events are also given. 
Our results show, in particular,
that for $\sin^2\theta_{23} \gtap 0.50$
$\sin^2\theta_{13} \gtap 0.01$ and
$\Delta m^2_{31} > 0$,
the effects of the Earth matter enhanced
$\nu_{\mu} \rightarrow \nu_e$ 
and $\nu_{e} \rightarrow \nu_{\mu}$
transitions of the atmospheric 
$\nu_{\mu}$ and $\nu_e$,
might be observable with the 
Super-Kamiokande detector.
Conversely, if the indicated effects 
are indeed observed
in the Super-Kamiokande experiment,
that would suggest in turn 
that $\sin^2\theta_{13} \gtap 0.01$, 
$\sin^2\theta_{23} \gtap 0.50$ and that 
the neutrino mass spectrum is with normal hierarchy,
$\Delta m^2_{31} > 0$. 

  Let us note finally that the Earth matter effects have
been widely investigated (for a recent detailed 
study see, e.g., ref. \cite{ConchaMal02} which contains also
a rather complete list of references to earlier
work on the subject). However, the study of 
the magnitude of the
Earth matter effects in the Nadir angle
distribution of the ratio of the multi-GeV
$\mu-$like and $e-$like events, 
measured in water-\v{C}erenkov 
detectors, performed here overlaps
very little with the earlier investigations.
 
\section{3-$\nu$ Oscillations of Atmospheric Neutrinos in the Earth}

\hskip 0.6truecm In the present Section we summarize the results on
the oscillations of atmospheric neutrinos 
crossing the Earth, which we use in our analysis.

\subsection{Preliminary Remarks}

\hskip 0.6truecm The $\nu_{\mu} \rightarrow \nu_{e}$
($\bar{\nu}_{\mu} \rightarrow \bar{\nu}_{e}$)
and $\nu_{e} \rightarrow \nu_{\mu (\tau)}$
($\bar{\nu}_{e} \rightarrow \bar{\nu}_{\mu (\tau)}$)
oscillations of atmospheric neutrinos should exist 
and their effects could be observable if
genuine three-flavour-neutrino  mixing takes place in vacuum,
i.e., if $\sin^22\theta_{13} \neq 0$, and
if $\sin^22\theta_{13}$ is sufficiently large 
\cite{SP3198} (see also, e.g., \cite{102,104,107,core}). 
Under the condition
$|\Delta m^2_{31}| \gg \Delta m^2_{21}$, which 
the neutrino mass squared differences 
determined by the existing atmospheric and 
solar neutrino and KamLAND data satisfy, 
the relevant three-neutrino 
$\nu_{\mu} \rightarrow \nu_{e}$ 
($\bar{\nu}_{\mu} \rightarrow \bar{\nu}_{e}$)
and $\nu_{e} \rightarrow \nu_{\mu (\tau)}$
($\bar{\nu}_{e} \rightarrow \bar{\nu}_{\mu (\tau)}$)
transition probabilities reduce effectively 
to a two-neutrino
transition probability \cite{3nuSP88}. 
with $\Delta m^2_{31}$  and $\sin^22\theta_{13} 
= 4|U_{e3}|^2(1 - |U_{e3}|^2)$ playing 
the role of the relevant
two-neutrino oscillation parameters.
Thus, searching for the effects of
$\nu_{\mu} \rightarrow \nu_{e}$ 
($\bar{\nu}_{\mu} \rightarrow \bar{\nu}_{e}$)
and $\nu_{e} \rightarrow \nu_{\mu (\tau)}$
($\bar{\nu}_{e} \rightarrow \bar{\nu}_{\mu (\tau)}$)
transitions of atmospheric neutrinos,
amplified by Earth matter effects, 
can provide unique information, 
in particular, about the magnitude of 
$\sin^2\theta_{13}$.

    As is not difficult to show using the results of \cite{3nuSP88},  
the 3-neutrino oscillation probabilities 
of interest for atmospheric $\nu_{e,\mu}$ having energy $E$
and crossing the Earth along a trajectory characterized by a 
Nadir angle $\theta_{n}$,
have the following form (see also, e.g., \cite{104,mantle}):
\begin{equation}
P_{3\nu}(\nu_{e} \rightarrow \nu_{e}) \cong 1 -
P_{2\nu}, 
\label{P3ee}
\end{equation}
\begin{equation}
P_{3\nu}(\nu_{e} \rightarrow \nu_{\mu}) \cong
P_{3\nu}(\nu_{\mu} \rightarrow \nu_{e}) \cong
s_{23}^2~P_{2\nu},
\label{P3emu}
\end{equation}
\begin{equation}
P_{3\nu}(\nu_{e} \rightarrow \nu_{\tau}) \cong
c_{23}^2~P_{2\nu},
\label{P3etau}
\end{equation}
\begin{equation}
P_{3\nu}(\nu_{\mu} \rightarrow \nu_{\mu}) \cong 1 -
s_{23}^4~P_{2\nu}
- 2c^2_{23}s^2_{23}~\left [ 1 -
Re~( e^{-i\kappa}
A_{2\nu}(\nu_{\tau} \rightarrow \nu_{\tau}))\right ] ,
\label{P3mumu}
\end{equation}
\begin{equation}
P_{3\nu}(\nu_{\mu} \rightarrow \nu_{\tau }) = 
1 - P_{3\nu}(\nu_{\mu} \rightarrow \nu_{\mu}) - 
P_{3\nu}(\nu_{\mu} \rightarrow \nu_{e}).
\label{P3mutau}
\end{equation}
\noindent  Here $P_{2\nu} \equiv 
P_{2\nu}(\Delta m^2_{31}, \theta_{13};E,\theta_{n})$ 
is the probability of two-neutrino 
oscillations in the Earth which coincides in form
with, e.g., the two-neutrino $\nu_{e} \rightarrow \nu_{\tau}$ transition
probability, $P_{2\nu}(\nu_{e} \rightarrow \nu_{\tau})$,
but describes $\nu_{e} \rightarrow \nu'_{\tau}$ 
transitions, where 
$\nu'_{\tau} = s_{23}\nu_{\mu} + c_{23} \nu_{\tau}$
\cite{3nuSP88}, and $\kappa$ and 
$A_{2\nu}(\nu_{\tau} \rightarrow \nu_{\tau}) \equiv A_{2\nu}$
are known phase and 
two-neutrino transition probability amplitude.
Analytic expressions for
$P_{2\nu}$, $\kappa$ and $A_{2\nu}$
will be given later.

  Using eqs. (\ref{P3ee}) - (\ref{P3mutau})
it is not difficult to convince oneself that  
the fluxes of atmospheric $\nu_{e,\mu}$ 
of energy $E$, which reach the detector after
crossing the Earth along a given trajectory  
specified by the value of $\theta_{n}$, 
$\Phi_{\nu_{e,\mu}}(E,\theta_{n})$, 
are given by the following expressions 
in the case of the three-neutrino oscillations  
under discussion \cite{102,104}:
\begin{equation}
\Phi_{\nu_e}(E,\theta_{n}) \cong 
\Phi^{0}_{\nu_e}~\left (  1 + 
  [s^2_{23}r - 1]~P_{2\nu}\right ),
\label{Phie}
\end{equation}
\begin{equation}
\Phi_{\nu_{\mu}}(E,\theta_{n}) \cong \Phi^{0}_{\nu_{\mu}} 
\left ( 1 +
 s^4_{23}~ [(s^2_{23}~r)^{-1} - 1]~P_{2\nu}
 - 2c^2_{23}s^2_{23}~\left [ 1 -
Re~( e^{-i\kappa}
A_{2\nu}(\nu_{\tau} \rightarrow \nu_{\tau})) \right ] \right )~, 
\label{Phimu}
\end{equation}
\noindent where $\Phi^{0}_{\nu_{e(\mu)}} = 
\Phi^{0}_{\nu_{e(\mu)}}(E,\theta_{n})$ is the
$\nu_{e(\mu)}$ flux in the absence of neutrino 
oscillations and
\vspace{-0.4cm}
\begin{equation}
r \equiv r(E,\theta_{n}) \equiv \frac{\Phi^{0}_{\nu_{\mu}}(E,\theta_{z})}
{\Phi^{0}_{\nu_{e}}(E,\theta_{z})}~~.
\label{r}
\end{equation}
\indent  The interpretation of the 
SK atmospheric 
neutrino data in terms of $\nu_{\mu} \rightarrow \nu_{\tau}$
oscillations requires the parameter 
$s^2_{23}$ to lie approximately in the interval
(0.30 - 0.70), with 0.5 being the statistically 
preferred value. For the predicted
ratio $r(E,\theta_{n})$ of the atmospheric 
$\nu_{\mu}$ and $\nu_e$ fluxes 
for i) the Earth core crossing and ii) only 
mantle crossing neutrinos, 
having trajectories for which
$0.4 \ltap \cos\theta_{n}\leq 1.0$, one has 
\cite{Honda,Bartol,Naumov}:
$r(E,\theta_{z}) \cong (2.0 - 2.5)$ for 
the neutrinos giving  
contribution to the sub-GeV
samples of Super-Kamiokande events, 
and $r(E,\theta_{n}) \cong (2.6 - 4.5)$ for those 
giving the main contribution to the multi-GeV samples.
If $s^2_{23} = 0.5$ and $r(E,\theta_{z}) \cong 2.0$,
we have $(s^2_{23}~r(E,\theta_{z}) - 1) \cong 0$ and
the possible effects of the 
$\nu_{\mu} \rightarrow \nu_{e}$ 
and $\nu_{e} \rightarrow \nu_{\mu (\tau)}$ 
transitions on the $\nu_e$ and $\nu_{\mu}$ 
fluxes, and correspondingly on 
the sub-GeV $e-$like sample of events, 
would be strongly suppressed even 
if these transitions were maximally
enhanced by the Earth matter effects.
For the multi-GeV neutrinos we have 
$(s^2_{23}~r(E,\theta_{z}) - 1) \gtap 0.3~(0.9)$ for 
$s^2_{23} = 0.5~(0.7)$.  
The factor $(s^2_{23}~r(E,\theta_{z}) - 1)$, 
for instance,
amplifies the effect of the 
$\nu_{\mu} \rightarrow \nu_{e}$ 
transitions in the $e-$like sample 
for $E \gtap (5 - 6)~{\rm GeV}$, for which
$r(E,\theta_{z}) \gtap 4$ \cite{Honda,Bartol,Naumov}. 

  The same conclusions are valid for the 
effects of oscillations on the fluxes of, and 
event rates due to, atmospheric antineutrinos:
$\bar{\nu}_e$ and $\bar{\nu}_{\mu}$.
Actually, the formulae for anti-neutrino 
fluxes and oscillation probabilities
are analogous to those for neutrinos: they can be obtained
formally from eqs. (\ref{P3ee}) - (\ref{r})
by replacing the neutrino related quantities - 
probabilities, $\kappa$,
$A_{2\nu}(\nu_{\tau} \rightarrow \nu_{\tau})$ 
and fluxes, with the 
corresponding quantities for antineutrinos:
$P_{2\nu}(\Delta m^2_{31}, \theta_{13};E,\theta_{n})
\rightarrow \bar{P}_{2\nu}(\Delta m^2_{31}, \theta_{13};E,\theta_{n})$,
$\kappa \rightarrow \bar{\kappa}$, 
$A_{2\nu}(\nu_{\tau} \rightarrow \nu_{\tau}) \rightarrow
A_{2\nu}(\bar{\nu}_{\tau} \rightarrow \bar{\nu}_{\tau}) 
\equiv \bar{A}_{2\nu}$,
$P_{3\nu}(\nu_{l} \rightarrow \nu_{l'}) \rightarrow
P_{3\nu}(\bar{\nu}_{l} \rightarrow \bar{\nu}_{l'})$,
$\Phi^{(0)}_{\nu_{e,\mu}}(E,\theta_{n}) \rightarrow 
\Phi^{(0)}_{\bar{\nu}_{e,\mu}}(E,\theta_{n})$ and
$r(E,\theta_{n}) \rightarrow \bar{r}(E,\theta_{n})$.

   Equations (\ref{P3ee}) - (\ref{P3mumu}), (\ref{Phie}) - (\ref{Phimu}) 
and the similar equations for
antineutrinos imply that in the case under study 
the effects of the $\nu_{\mu} \rightarrow \nu_{e}$,
$\bar{\nu}_{\mu} \rightarrow \bar{\nu}_{e}$, 
and $\nu_{e} \rightarrow \nu_{\mu (\tau)}$,
$\bar{\nu}_{e} \rightarrow \bar{\nu}_{\mu (\tau)}$,
oscillations 
i) increase with the increase of $s^2_{23}$ and are maximal
for the largest allowed value of $s^2_{23}$,
ii) should be considerably larger in the multi-GeV 
samples of events than in the sub-GeV samples,
iii) in the case of the multi-GeV samples,
they lead to an increase of the
rate of  $e$-like 
events and to a slight decrease of the 
$\mu-$like event rate.
This discussion suggests that the quantity
most sensitive to the effects of 
the oscillations of interest
should be the ratio of the 
$\mu-$like and $e-$like 
multi-GeV events (or event rates), 
$N_{\mu}/N_{e}$.

  The magnitude of the effects we are interested in 
depends also on the 2-neutrino oscillation probabilities,
$P_{2\nu}(\Delta m^2_{31}, \theta_{13};E,\theta_{n})$ and
$\bar{P}_{2\nu}(\Delta m^2_{31}, \theta_{13};E,\theta_{n})$.
In the case of oscillations in vacuum we have
$P_{2\nu}(\Delta m^2_{31}, \theta_{13};E,\theta_{n}) =
\bar{P}_{2\nu}(\Delta m^2_{31}, \theta_{13};E,\theta_{n})
\sim \sin^22\theta_{13}$. Given the existing limits on
$\sin^22\theta_{13}$, the 
probabilities $P_{2\nu}$ and
$\bar{P}_{2\nu}$ cannot be large if the oscillations 
take place in vacuum. However, $P_{2\nu}$ or
$\bar{P}_{2\nu}$ can be strongly enhanced by the
Earth matter effects. The latter depend on the
Earth density profile and we will discuss it next briefly.
\subsection{The Earth Model and the Two-Layer Density Approximation}

\hskip 0.6truecm As is well-known, the Earth density distribution
in the existing Earth models is assumed to be 
spherically symmetric
\footnote{Let us note that 
because of the approximate 
spherical symmetry of the Earth, 
a given neutrino trajectory through 
the Earth is completely 
specified by its Nadir angle.}
and there are two major density structures - 
the core and the mantle, and 
a certain number of substructures (shells or layers).
The core radius and the depth of the mantle
are known with a rather good precision 
and these data are incorporated
in the Earth models.
According to the Stacey 1977 and the more recent PREM models
\cite{Stacey:1977,PREM81}, which are widely used in the 
calculations of the probabilities of neutrino oscillations
in the Earth, the core has a radius $R_c = 3485.7~$km,
the Earth mantle depth is approximately $R_{man} = 2885.3~$km,
and the Earth radius is $R_{\oplus} = 6371~$km.  
The mean values of the matter densities of the core and of the mantle 
read, respectively: $\bar{\rho}_c \cong 11.5~{\rm g/cm^3}$ and 
$\bar{\rho}_{man} \cong 4.5~{\rm g/cm^3}$.

   All the interesting features of the 
atmospheric neutrino oscillations in the Earth
can be understood quantitatively 
in the framework of the two-layer 
model of the Earth density
distribution \cite{SP3198}.
The density profile of the Earth in the two-layer model 
is assumed to consist of two structures - 
the mantle and the core, 
having different densities, $\rho_{man}$ and $\rho_{c}$, and different
electron fraction numbers, $Y_e^{man}$ and $Y_e^{c}$, none of which however 
vary within a given structure.
The densities $\rho_{man}$ and $\rho_{c}$
in the case of interest should be considered 
as mean effective densities
along the neutrino trajectories, and they vary somewhat with
the change of the trajectory
\cite{SP3198}: 
$\rho_{man} = \bar{\rho}_{man}$ and 
$\rho_c = \bar{\rho}_{c}$.
In the PREM and Stacey models one has
for $\cos \theta_{n} \gtap 0.4$:
$\bar{\rho}_{man} \cong (4 - 5)~ {\rm g/cm^3}$ and 
$\bar{\rho}_{c} \cong (11 - 12)~ {\rm g/cm^3}$.
For the electron fraction numbers in 
the mantle and in the core 
\footnote{The electron fraction number is given by 
$Y_e = N_e/(N_p + N_n) = N_p/(N_p + N_n)$, where 
$N_e$, $N_p$ and $N_n$ are the electron, proton and neutron
number densities in matter, respectively.}
one can use the standard values \cite{Art2} 
$Y_e^{man} = 0.49$ and $Y_e^{c} = 0.467$.
Numerical calculations show 
\cite{MP98:2layers} (see also \cite{SP3198,3nuKP88}) 
that, e.g., the 
$\nu_{e} \rightarrow \nu_{\mu}$
oscillation probability of interest,
calculated within 
the two-layer model of the Earth with $\bar{\rho}_{man}$
and $\bar{\rho}_{c}$ for a given neutrino trajectory
determined using the PREM (or the Stacey) model,  
reproduces with a remarkably high precision 
the corresponding probability calculated 
by solving numerically the 
relevant system of evolution equations
with the much more sophisticated Earth density profile
of the PREM (or Stacey) model. 
  
   We give below the expressions for the probability 
$P_{2\nu}(E,\theta_{n}; \Delta m^2_{31}, \theta_{13})$, 
for the phase $\kappa$ and 
for the amplitude $ A_{2\nu}(\nu_{\tau} \rightarrow \nu_{\tau})$
in the two-layer approximation of the Earth density distribution
and in the general case of neutrinos crossing the Earth core.
The expression for $P_{2\nu}(E,\theta_{n}; 
\Delta m^2_{31}, \theta_{13})$, as we have already indicated,
coincides with that for the 
probability of the two-neutrino $\nu_{\mu} \rightarrow \nu_{e}$ 
($\nu_{e} \rightarrow \nu_{\mu (\tau)}$) transitions
and has the form \cite{SP3198}
\footnote{The expression for $P_{2\nu}$ (eq.  
(\ref{P2nu})) can be obtained 
from the expression for the probability $P_{e2} = 
P(\nu_2 \rightarrow \nu_e)$ given in eq. (7) in \cite{SP3198}
by formally setting $\theta = 0$ while keeping 
$\theta_{m}' \neq 0$ and 
$\theta_{m}'' \neq 0$.}~:
$$P_{2\nu}(E,\theta_{z};\Delta m^2_{31}, \theta_{13}) 
=  {1\over {2}} 
\left [1 - \cos \Delta E''X'' \right ] \sin^2 2\theta''_{m} ~~~~~~~~~~
~~~~~~~~~~~~~~~~~~~~~~~~~~~~~~~~$$
$$~~~~ +~ {1\over {4}} \left [1 - \cos \Delta E''X'' \right ]
\left [1 - \cos \Delta E'X' \right ] 
\left [ \sin^2(2\theta''_{m} - 4\theta'_{m}) - 
 \sin^2 2\theta''_{m} \right ]~~~~~~~~~~~~~$$
$$~~~~ - ~{1\over {4}} \left [1 - \cos \Delta E''X'' \right ]
\left [1 - \cos 2\Delta E'X' \right ] \sin^2 2\theta'_{m}  
\cos^2(2\theta''_{m} - 2\theta'_{m})~~~~~~~~~~~~~~~~~$$
$$~+~{1\over {4}} \left [1 + \cos \Delta E''X'' \right ]
\left [1 - \cos 2\Delta E'X' \right ]~\sin^2 2\theta'_{m}~~~~~~~~
~~~~~~~~~~~~~~~~~~~~~~~~~~~ $$
$$+ ~{1\over {2}} \sin \Delta E''X'' ~\sin 2\Delta E'X'  
~\sin^2 2\theta'_{m} \cos(2\theta''_{m} - 2\theta'_{m})
~~~~~~~~~~~~~~~~~~~~~~~~~~$$
\begin{equation}
~~~~~ +~{1\over {4}} \left [ \cos (\Delta E'X' - \Delta E''X'') 
- \cos (\Delta E'X' + \Delta E''X'') \right ]
 \sin 4\theta'_{m} 
 \sin (2\theta''_{m} - 2\theta'_m).  
\label{P2nu}
\end{equation}
\noindent Here
\begin{equation}
\Delta E'~(\Delta E'') = \\ \nonumber
\frac{\Delta m^2_{31}}{2E}
\sqrt{\left( 1 - \frac{\bar{\rho}_{man~(c)}}{\rho^{res}_{man~(c)}}
\right)^2 \cos^22\theta_{13} 
+ \sin^22\theta_{13} }~,~
\label{DE}
\end{equation}
\noindent is the difference between the energies of the two energy-
eigenstate neutrinos in the mantle (core),
$\theta'_{m}$ and  $\theta''_{m}$ are 
the mixing angles in matter in the
mantle and in the core, respectively,
\begin{equation}
\sin^22\theta'_{m}~(\sin^22\theta''_{m}) = \frac{\sin^22\theta_{13}}
{(1 - \frac{\bar{\rho}_{man(c)}}{\rho^{res}_{man(c)}})^2 
\cos^22\theta_{13} + 
\sin^22\theta_{13} },
\label{thetam}
\end{equation}
\noindent $X'$ is half of the distance the neutrino
travels in the mantle and $X''$ is the length of 
the path of the neutrino in the core,
$\rho^{res}_{man}$ and $\rho^{res}_{c}$
are the resonance densities in the mantle and in the
core, and  $\bar{\rho}_{man}$ and $\bar{\rho}_{c}$
are the mean densities along the neutrino trajectory
in the mantle and in the core 
For a neutrino trajectory which is 
specified by a given Nadir angle $\theta_n$
we have:
\begin{equation}
X' = R_{\oplus}\cos \theta_n -  \sqrt{R^{2}_{c} - 
R^2_{\oplus}\sin^2\theta_n},~~
X'' = 2\sqrt{R^{2}_{c} - R^2_{\oplus}\sin^2\theta_n},~
\end{equation}
\noindent where $R_{\oplus} = 6371~$km is 
the Earth radius (in the PREM \cite{PREM81} and 
Stacey \cite{Stacey:1977} models)
and $R_c = 3485.7~$km  is the core radius. 
The neutrinos cross the Earth core on the 
way to the detector for $\theta_{n} \ltap 33.17^{o}$.

   The  resonance densities in the mantle and in the
core can be obtained from the expressions 
\begin{equation}
\rho^{res} = \frac{\Delta m^2_{31} \cos2\theta_{13}}{2E\sqrt{2}
G_F Y_e}~m_{N}~,
\label{rhores}
\end{equation}
\noindent $m_{N}$ being the nucleon mass,
by using the specific values of $Y_e$ in the mantle and in the core.
We have $\rho^{res}_{man}\neq \rho^{res}_{c}$  
because $Y_e^{c} = 0.467$ and $Y_e^{man} = 0.49$ \cite{Art2}
(see also \cite{EarthCORE}). Obviously, 
$Y_e^{c}\rho^{res}_{c} = Y_e^{man}\rho^{res}_{man}$.

  The phase $\kappa$ and the probability 
amplitude $A_{2\nu}$
which appear in eq. (\ref{Phimu}) 
for the flux of atmospheric $\nu_{\mu}$
in the case of three-flavour neutrino mixing and hierarchy
between the neutrino mass squared differences and therefore 
can play important role in the interpretation of the, 
e.g., Super-Kamiokande
atmospheric neutrino data, have the following form 
in the two-layer model of the Earth density distribution 
\cite{SP3198,102,104}:
\begin{equation}
\kappa \cong {1\over {2}} [ {\Delta m^2_{31}\over{2E}}~X +
\sqrt{2} G_F {1\over{m_N}}(X''Y_e^{c}\bar{\rho}_{c} +
2X'Y_e^{man}\bar{\rho}_{man}) -
2\Delta E'X' - \Delta E''X'']
- {\Delta m^2_{21}\over{2E}}X\cos2\theta_{12},
\label{kappa}
\end{equation}
\begin{equation}
A_{2\nu}(\nu_{\tau} \rightarrow \nu_{\tau}) = 1 +~
\left( e^{-i2\Delta E'X'} - 1 \right) \left[ 1 +
\left(e^{-i\Delta E''X''} - 1 \right)
\cos^2(\theta'_{m} - \theta''_{m}) \right ]
\cos^2\theta'_{m}~
\label{A2nu}
\end{equation}
\vspace*{-0.6cm}
$$ +~ \left( e^{-i\Delta E''X''} - 1 \right)
\cos^2\theta''_{m}~+~ {1\over {2}} ~\left( e^{-i\Delta E''X''} - 1 \right )~
\left( e^{-i\Delta E'X'} - 1 \right) \sin(2\theta'_{m} - 2\theta''_{m})
\sin2\theta'_{m}~,$$ 
\noindent where 
\footnote{One can get the expression for the amplitude 
$A_{2\nu}(\nu_{\tau} \rightarrow \nu_{\tau})$
from eq. (1) in ref. \cite{SP3198} by formally setting 
$\theta = \pi/2$ while keeping $\theta'_m$ and $\theta''_m$ 
arbitrary, and then interchanging $\sin\theta'_m$ ($\sin \theta''_m$) and
$\cos\theta'_m$ ($\cos\theta''_m$).}
$X = X'' + 2X'$. 

   The expressions for 
$\bar{P}_{2\nu}$, $\bar{\kappa}$ and 
$A_{2\nu}(\bar{\nu}_{\tau} \rightarrow \bar{\nu}_{\tau})$
can be obtained from the corresponding 
expressions for neutrinos by replacing 
$\rho_{man,c}$ with (-$\rho_{man,c}$)
in eqs. (\ref{DE}) and (\ref{thetam}).

\subsection{Oscillations in the Earth Mantle}

\hskip 0.6truecm 
In the two-layer model, the oscillations of atmospheric 
neutrinos crossing only the Earth mantle (but not 
the Earth core), correspond to oscillations in matter with
constant density. The relevant expressions for 
$P_{2\nu}$, $\kappa$ and 
$A_{2\nu}(\nu_{\tau} \rightarrow \nu_{\tau})$
follow from eqs. (\ref{P2nu}), (\ref{kappa}) and  
(\ref{A2nu}) by setting $X'' = 0$ and using 
$X' = R_{\oplus}\cos \theta_n$. The expressions 
for $P_{2\nu}$ ($\bar{P}_{2\nu}$) and 
$A_{2\nu}$ ($\bar{A}_{2\nu}$)
have the standard well-known form.

   In the case under study $\theta_{13}$ plays 
the role of a two-neutrino vacuum mixing angle 
in the probabilities $P_{2\nu}$ and
$\bar{P}_{2\nu}$. Since $\sin^2\theta_{13} < 0.05$,
we have $\cos2\theta_{13} > 0$. Consequently, 
the Earth matter effects can 
resonantly enhance $P_{2\nu}$ 
for $\Delta m^2_{31} > 0$ 
and $\bar{P}_{2\nu}$
if $\Delta m^2_{31} < 0$ \cite{mantle}. 
Due to the difference of cross sections 
for neutrinos and antineutrinos,
approximately 2/3 of the total rate 
of the $\mu-$like and $e-$like multi-GeV atmospheric neutrino 
events in the SK (and in any other water-\v{C}erenkov) detector, i.e.,
$\sim 2N_{\mu}/3$ and $\sim 2N_e/3$,
are due to neutrinos $\nu_{\mu}$ and $\nu_e$, respectively,
while the remaining $\sim 1/3$ of the multi-GeV 
event rates, i.e., $\sim N_{\mu}/3$ and $\sim N_e/3$,
are produced by antineutrinos
$\bar{\nu}_{\mu}$ and $\bar{\nu}_e$.
This implies that the Earth matter effects 
in the multi-GeV samples of $\mu-$like and $e-$like 
events will be larger if $\Delta m^2_{31} > 0$, i.e., 
if the neutrino mass spectrum 
is with normal hierarchy, than if $\Delta m^2_{31} < 0$ and 
the spectrum is with inverted hierarchy.
Thus, the ratio $N_{\mu}/N_e$ of the 
multi-GeV $\mu-$like and $e-$like event rates
measured in the SK experiment
is sensitive, in principle, to the type of the 
neutrino mass spectrum.

     Consider for definiteness the case of $\Delta m^2_{31} > 0$.
It follows from eqs. (\ref{Phie}) and (\ref{Phimu})
that the oscillation effects of interest 
will be maximal if $P_{2\nu} \cong 1$.
The latter is possible provided 
i) the well-known 
resonance condition \cite{BPPW80,MS85}, leading  
to $\sin^22\theta'_m \cong 1$, is fulfilled, and  
ii) $\cos 2\Delta E' X' \cong -1$.
Given the values of  
$\bar{\rho}_{man}$ and $Y_e^{man}$,
the first condition determines the
neutrino energy at which 
$P_{2\nu}$ can be enhanced:
\begin{equation}
E_{res} \cong 6.6 \times \Delta m^2_{31}[10^{-3}~{\rm eV^2}]~ 
( {\rm \bar{N}_e^{man}[N_A cm^{-3}]})^{-1} \cos2\theta_{13}~{\rm GeV},
\label{Eres}
\end{equation}
\noindent where $\Delta m^2_{31}[10^{-3}~{\rm eV^2}]$
is the value of $\Delta m^2_{31}$ in units of
$10^{-3}~{\rm eV^2}$ and
${\rm \bar{N}_e^{man}[N_A~cm^{-3}}]$ is the electron
number density, ${\rm \bar{N}_e^{man} = 
Y_e^{man} \bar{\rho}_{man}/m_N}$, in units of
${\rm N_A cm^{-3}}$,
$N_A$ being the Avogadro number.
If the first condition is satisfied, the second 
determines the length of the path of the neutrinos 
in the mantle for which one can have
$P_{2\nu} \cong 1$:
\begin{equation}
2X'(\Delta E')_{res} \cong 1.2\pi~\tan2\theta_{13}~  
{\rm \bar{N}_e^{man}[N_A cm^{-3}]}~2X'[10^{4}~{\rm km}]~,
\label{Xman}
\end{equation}
\noindent where $X'$ is in units of $10^{4}~{\rm km}$.
Taking $\Delta m^2_{31} \cong (2.1 - 3.3)
\times 10^{-3}~{\rm eV^2}$,
${\rm \bar{N}_e^{man} \cong 2~{\rm N_A cm^{-3}}}$
and $\cos2\theta_{13} \cong 1$ one finds 
from eq. (\ref{Eres}): $E_{res} \cong (7 - 11)~{\rm GeV}$.
The width of the resonance in $E$, 2$\delta E$,
is determined, as is well-known, 
by $\tan2\theta_{13}$:
$\delta E/E_{res} \sim \tan 2\theta_{13}$. 
For $\sin^2\theta_{13} \sim (0.01 - 0.05)$,
the resonance is relatively wide in the neutrino energy:
$\delta E/E_{res} \cong (0.27 - 0.40)$.
Equation (\ref{Xman}) implies that 
for $\sin^2\theta_{13} = 0.05~(0.025)$ 
and $\bar{N}_e^{man} \cong 2~{\rm N_A cm^{-3}}$,
one can have $P_{2\nu} \cong 1$
only if $2X' \cong 8000~(10000)~{\rm km}$.

   It follows from the above simple analysis 
\cite{mantle}
that the Earth matter effects can amplify $P_{2\nu}$
significantly when the neutrinos cross 
only the mantle i) for 
$E \sim (5 - 10)$ GeV, i.e., in the
multi-GeV range of neutrino energies,
and ii) only for sufficiently long 
neutrino paths in the mantle, i.e., for
$\cos\theta_n \gtap 0.4$.
The magnitude of the matter effects in the ratio
$N_{\mu}/N_e$ of interest 
increases with increasing 
$\sin^2\theta_{13}$.

 The same results, eqs. (\ref{Eres}) and (\ref{Xman}),
and conclusions
are valid for the antineutrino oscillation probability
$\bar{P}_{2\nu}$ in the case of 
$\Delta m^2_{31} < 0$. As a consequence, the ideal
situation for distinguishing the type of mass hierarchy
would be a detector with charge discrimination. 
\subsection{Oscillations of Atmospheric Neutrinos
Crossing the Earth Core}

\hskip 0.6truecm  In this case $P_{2\nu}$, $\kappa$ and 
$A_{2\nu}$ are given by eqs. (\ref{P2nu}), 
(\ref{kappa}) and  (\ref{A2nu}). 
In the discussion which follows we will 
assume for concretness that 
$\Delta m^2_{31} > 0$, 
consider the probability
$P_{2\nu}$ and the transitions 
of neutrinos.
If $\Delta m^2_{31} < 0$, 
the results we will briefly discuss
below will be valid for the probability 
$\bar{P}_{2\nu}$ and the
transitions of antineutrinos.

  For $\sin^2\theta_{13} < 0.05$ and
$\Delta m^2_{31} > 0$,
we can have $P_{2\nu} \cong 1$ 
{\it only} due to the effect of maximal constructive 
interference between the amplitudes of the 
the $\nu_{e} \rightarrow \nu'_{\tau}$
transitions in the Earth mantle and in the 
Earth core \cite{SP3198,106,107}.
The effect differs from the MSW one \cite{SP3198}.
The  {\it mantle-core enhancement effect} 
is caused by the existence 
(for a given neutrino trajectory
through the Earth core) of 
{\it points of resonance-like 
total neutrino conversion}, 
$P_{2\nu} = 1$,
in the corresponding space 
of neutrino oscillation 
parameters \cite{106,107}. 
The location of these points 
determines the regions
where $P_{2\nu}$ 
is large, $P_{2\nu} \gtap 0.5$. 
These regions vary slowly with the Nadir angle,
they are remarkably wide in the Nadir angle
and are rather wide in the neutrino energy \cite{107},
so that the transitions of interest produce noticeable 
effects in the ratio $N_{\mu}/N_e$:
we have, e.g.,  $\delta E/E \cong 0.3$  
for the values of $\sin^2\theta_{13}$
of interest.
 
 The resonance-like total neutrino conversion 
due to the mantle-core enhancement effect
takes place for a given neutrino trajectory 
through the Earth core if the following two 
conditions are satisfied \cite{106,107}:
\begin{equation}
\tan \frac{1}{2}\Delta E' X' = 
\pm \sqrt{\frac{-\cos2\theta_m''}{\cos(2\theta_m'' - 4\theta_m')}},~~
\tan \frac{1}{2}\Delta E'' X'' = 
\pm \frac{\cos2\theta_m'}
{\sqrt{-\cos2\theta_m''\cos(2\theta_m'' - 4\theta_m')}},~~
\label{NOLR1}
\end{equation}
where the signs are correlated and 
$\cos2\theta_m''\cos(2\theta_m'' - 4\theta_m') \leq 0$.
As was shown in \cite{107},
conditions (\ref{NOLR1})
are fulfilled for the 
$\nu_{\mu} \rightarrow \nu_{e}$ and 
$\nu_{e} \rightarrow \nu_{\mu (\tau)}$ 
transitions of the Earth core-crossing atmospheric 
neutrinos. A rather complete set of values of 
$\Delta m^2_{31}/E$ and $\sin^22\theta_{13}$
for which both conditions in (\ref{NOLR1}) 
hold and $P_{2\nu} = 1$
for neutrino trajectories 
with Nadir angle $\theta_n = 0;~13^0;~23^0;~30^0$
was also found in \cite{107}. 

  For $\sin^2\theta_{13} < 0.05$,
there are two sets of values of 
$\Delta m_{31}^2$ and $\sin^2\theta_{13}$
for which eq. (\ref{NOLR1})
is fulfilled and $P_{2\nu} = 1$.
These two solutions of eq. (\ref{NOLR1})
occur for, e.g., $\theta_n = 0;~13^0;23^0$,
at  1) $\sin^22\theta_{13} = 0.034;~0.039;~0.051$, 
$\Delta m_{31}^2/E = 7.2;~7.0;~6.5~\times10^{-7}~{\rm eV^2/MeV}$,
and at 2) $\sin^22\theta_{13} = 0.15;~0.17;~0.22$, 
$\Delta m_{31}^2/E = 4.8;~4.5;~3.8~ \times 10^{-7}~{\rm eV^2/MeV}$
(see Table 2 in ref. \cite{107}).
The first solution corresponds to
$\cos 2\Delta E' X' = -1$,
$\cos \Delta E'' X'' = -1$ and
\footnote{The term
``neutrino oscillation length resonance'' (NOLR) 
was used in \cite{SP3198}
to denote the mantle-core 
enhancement effect
in this case.}
and $\sin^2(2\theta_m'' - 4\theta_m') = 1$. 
For $\Delta m^2 = 3\times 10^{-3}~{\rm eV^2}$, the 
total neutrino conversion
occurs in the case of the first solution
at $E \cong (4.2 - 4.7)~{\rm GeV}$.
The atmospheric $\nu_{e}$ and  $\nu_{\mu}$
with these energies 
contribute to the multi-GeV samples of 
$e-$like and $\mu-$like events
in the SK experiment.
The values of $\sin^22\theta_{13}$ 
at which the second
solution takes place are marginally allowed.
If $\Delta m^2 = 3\times 10^{-3}~{\rm eV^2}$,
one has $P_{2\nu} = 1$
for this solution 
for a given $\theta_n$  in the interval
$0 \ltap \theta_n \ltap 23^0$
at $E$ lying in the interval
$E \cong (6.3 - 8.0)~{\rm GeV}$.

   The above discussion suggests, 
in particular, that the 
effects of the mantle-core 
(NOLR) enhancement of $P_{2\nu}$ 
(or $\bar{P}_{2\nu}$) in the ratios
$N_{\mu}/N_e$ and $N^{3\nu}_{e}/N^{0}_e$
increase rapidly with $\sin^2\theta_{13}$
as long as $\sin^2\theta_{13}\ltap 0.01$,
and should exhibit a rather weak dependence on
$\sin^2\theta_{13}$ for
$0.01 \ltap \sin^2\theta_{13} < 0.05$.
If 3-neutrino oscillations of 
atmospheric neutrinos take place,
the magnitude of the matter effects 
in the multi-GeV $e-$like 
and $\mu-$like event samples, 
produced by neutrinos 
crossing the Earth core, 
should be larger than 
in the event samples
due to neutrinos
crossing only the Earth mantle (but not the core).
This is a consequence of the fact 
that in the energy range of interest
the atmospheric neutrino fluxes 
decrease rather rapidly with
energy - approximately as $E^{-2.7}$,
while the neutrino interaction 
cross section rises only linearly 
with $E$, and that the maximum
of $P_{2\nu}$ (or $\bar{P}_{2\nu}$)
due to the NOLR takes place 
at approximately two times smaller
energies than that
due to the MSW effect
for neutrinos crossing only the
Earth mantle (e.g., at $E \cong (4.2 - 4.7)~{\rm GeV}$
and $E \cong 10~{\rm GeV}$, respectively,
for $\Delta m^2 = 3\times 10^{-3}~{\rm eV^2}$).

\section{Results}

\hskip 0.6truecm The results of our analysis 
are summarized graphically in Figs. 1 - 5. We have 
used in the calculations the 
predictions for the Nadir angle and 
energy distributions of the
atmospheric neutrino fluxes given in \cite{Naumov}. 
In this analysis,
we only consider Deep Inelastic Scattering (DIS) 
cross sections and we
make use of the GRV94 parton distributions 
given in \cite{GRV94}.

\begin{figure}
\includegraphics[height=18.2cm]{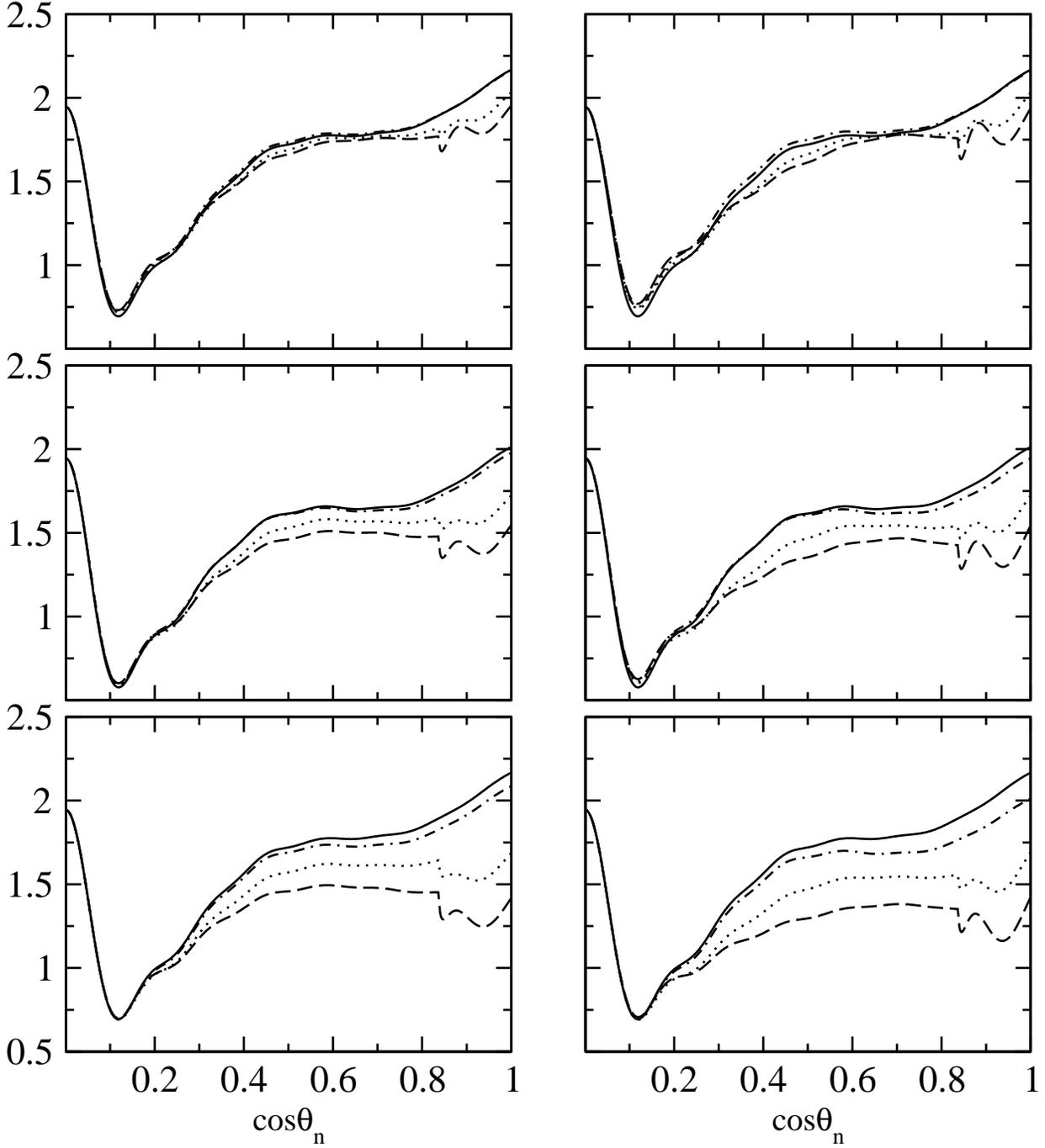}
\caption{\label{nadir} 
 The dependence on $\cos\theta_n$ 
of the ratios of the multi-GeV
$\mu-$ and $e-$ like events 
(or event rates), integrated
over the neutrino energy 
in the interval $E = (2.0 - 10.0)$ GeV, 
in the cases 
i) of two-neutrino
$\nu_{\mu} \rightarrow \nu_{\tau}$ 
and $\bar{\nu}_{\mu} \rightarrow \bar{\nu}_{\tau}$
oscillations in vacuum and
no $\nu_e$ and $\bar{\nu}_e$ oscillations,
$N^{2\nu}_{\mu}/N^{0}_{e}$ (solid lines),
ii) three-neutrino
oscillations in vacuum of 
$\nu_{\mu}$, $\bar{\nu}_{\mu}$,
$\nu_e$ and $\bar{\nu}_e$,
$(N^{3\nu}_{\mu}/N^{3\nu}_{e})_{vac}$ (dash-dotted lines),
iii) three-neutrino
oscillations of 
$\nu_{\mu}$, $\bar{\nu}_{\mu}$
$\nu_e$ and $\bar{\nu}_e$ in the Earth
and neutrino mass spectrum
with normal hierarchy 
$(N^{3\nu}_{\mu}/N^{3\nu}_{e})_{\rm NH}$ (dashed lines),
or with inverted hierarchy,
$(N^{3\nu}_{\mu}/N^{3\nu}_{e})_{\rm IH}$ (dotted lines).
The results shown are
for $|\Delta m^2_{31}| = 3\times 10^{-3}~{\rm eV^2}$,
$\sin^2\theta_{23} = 0.36~(\rm upper~panels);~0.50
~(\rm middle~panels);~0.64~(\rm lower~panels)$,
and $\sin^22\theta_{13} = 0.05~(\rm left~panels);
~0.10 (\rm right~panels)$.
}
\vspace{-1cm}   
\end{figure}

 The predicted dependences
on $\cos\theta_n$ 
of the ratios of the multi-GeV
$\mu-$ and $e-$ like events
(or event rates), integrated
over the neutrino energy 
from the interval $E = (2.0 - 10.0)$ GeV,
in the case 
i) of two-neutrino
$\nu_{\mu} \rightarrow \nu_{\tau}$ 
and $\bar{\nu}_{\mu} \rightarrow \bar{\nu}_{\tau}$
oscillations in vacuum and
no $\nu_e$ and $\bar{\nu}_e$ oscillations,
$N^{2\nu}_{\mu}/N^{0}_{e}$,
ii) three-neutrino
oscillations in vacuum of 
$\nu_{\mu}$, $\bar{\nu}_{\mu}$,
$\nu_e$ and $\bar{\nu}_e$,
$(N^{3\nu}_{\mu}/N^{3\nu}_{e})_{vac}$,
iii) three-neutrino
oscillations of 
$\nu_{\mu}$, $\bar{\nu}_{\mu}$,
$\nu_e$ and $\bar{\nu}_e$ in the Earth
in the cases of neutrino mass spectrum
with normal hierarchy 
$(N^{3\nu}_{\mu}/N^{3\nu}_{e})_{\rm NH}$,
and with inverted hierarchy,
$(N^{3\nu}_{\mu}/N^{3\nu}_{e})_{\rm IH}$,
for $\sin^2\theta_{23} = 0.64;~0.50;~0.36$,
$\sin^22\theta_{13} = 0.05;~0.10$ and
$|\Delta m^2_{31}| = 3\times 10^{-3}~{\rm eV^2}$
are shown in Fig. 1. 

   As $\cos\theta_n$ increases from 0 to $\sim 0.2$,
the neutrino path length in the Earth increases 
from 0 to $2X' = 2R_{\oplus} \cos\theta_n \cong  2550$ km.
The $\nu_\mu \rightarrow \nu_{\tau}$
($\bar{\nu}_\mu \rightarrow \bar{\nu}_{\tau}$)
oscillations, which for $\cos\theta_n \ltap 0.2$
proceed in the Earth essentially as in vacuum,
fully develop. For $|\Delta m^2_{31}| = 
3\times 10^{-3}~{\rm eV^2}$
and $E = 3$ GeV, for example, 
the maximum of the 
$\nu_\mu \rightarrow \nu_{\tau}$
($\bar{\nu}_\mu \rightarrow \bar{\nu}_{\tau}$)
oscillation probability
occurs for $\cos\theta_n \cong 0.1$, or 
$2X' \cong 1270$ km.
For $\cos\theta_n \ltap 0.2$ and
$|\Delta m^2_{31}| = 3\times 10^{-3}~{\rm eV^2}$,
the oscillations involving the atmospheric 
$\nu_e$ ($\nu_e \rightarrow \nu_{\mu,\tau}$,
$\nu_\mu \rightarrow \nu_{e}$) 
and $\bar{\nu}_e$ ($\bar{\nu}_e \rightarrow \bar{\nu}_{\mu,\tau}$,
$\bar{\nu}_\mu \rightarrow \bar{\nu}_{e}$)
with energies in the multi-GeV range,
$E \sim (2.0 - 10.0)$ GeV, are suppressed.
If $\Delta m^2_{31} > 0$, for instance,
the Earth matter effects suppress
the antineutrino oscillation probability
$\bar{P}_{2\nu}$, but can enhance 
the neutrino mixing in matter, or
$\sin^22\theta'_m$. However, 
since the neutrino path in the Earth mantle
is relatively short one has
$2X'\Delta E' \ll 1$, and correspondingly
$P_{2\nu} \ll 1$.
Thus, for  $\cos\theta_n \ltap 0.2$,
all four types of ratios we consider, 
$N^{2\nu}_{\mu}/N^{0}_{e}$,
$(N^{3\nu}_{\mu}/N^{3\nu}_{e})_{vac}$,
$(N^{3\nu}_{\mu}/N^{3\nu}_{e})_{\rm NH}$
and $(N^{3\nu}_{\mu}/N^{3\nu}_{e})_{\rm IH}$,
practically coincide and exhibit 
the same dependence on $\cos\theta_n$:
they decrease as $\cos\theta_n$
increases from 0, reaching a minimum
at $\cos\theta_n \cong 0.1$, and begin
to rise as $\cos\theta_n$ increases 
further. This behavior is clearly seen
in Figs. 1 and 2.

  At $\cos\theta_n \gtap 0.4$,
the Earth matter effects in the oscillations
of the atmospheric $\nu_{\mu}$, $\bar{\nu}_{\mu}$
$\nu_e$ and $\bar{\nu}_e$,
can generate noticeable differences
between $N^{2\nu}_{\mu}/N^{0}_{e}$ (or
$(N^{3\nu}_{\mu}/N^{3\nu}_{e})_{vac}$)
and $(N^{3\nu}_{\mu}/N^{3\nu}_{e})_{\rm NH(IH)}$, 
as well as between
$(N^{3\nu}_{\mu}/N^{3\nu}_{e})_{\rm NH}$ and
$(N^{3\nu}_{\mu}/N^{3\nu}_{e})_{\rm IH}$.
For $\sin^2\theta_{23} = 0.36$ and
$\sin^22\theta_{13} \ltap 0.05$
(upper left panel in Fig. 1),
we have at $\cos\theta_n \ltap 0.8$
(neutrinos crossing only the Earth mantle):
$(N^{3\nu}_{\mu}/N^{3\nu}_{e})_{\rm NH} \cong
(N^{3\nu}_{\mu}/N^{3\nu}_{e})_{\rm IH} \cong
 N^{2\nu}_{\mu}/N^{0}_{e} \cong
 (N^{3\nu}_{\mu}/N^{3\nu}_{e})_{vac}$.
For the Earth-core-crossing 
atmospheric neutrinos, $\cos\theta_n \gtap 0.84$,
the mantle-core interference effect (or NOLR) suppresses
the ratios $(N^{3\nu}_{\mu}/N^{3\nu}_{e})_{\rm NH,~IH}$
with respect to $N^{2\nu}_{\mu}/N^{0}_{e}$
(or $(N^{3\nu}_{\mu}/N^{3\nu}_{e})_{vac}$):
at $\sin^22\theta_{13} = 0.10$ the relative 
averaged difference between
$N^{2\nu}_{\mu}/N^{0}_{e}$ and
$(N^{3\nu}_{\mu}/N^{3\nu}_{e})_{\rm NH}$ is 
\footnote{The relative 
difference of, e.g, 
$N^{2\nu}_{\mu}/N^{0}_{e}$ and
$(N^{3\nu}_{\mu}/N^{3\nu}_{e})_{\rm NH}$
is defined as
$(1 - (N^{3\nu}_{\mu}/N^{3\nu}_{e})_{\rm NH}/
(N^{2\nu}_{\mu}/N^{0}_{e}))$.}
$11\%$,
while the difference between
$(N^{3\nu}_{\mu}/N^{3\nu}_{e})_{\rm NH}$ and
$(N^{3\nu}_{\mu}/N^{3\nu}_{e})_{\rm IH}$
is rather small (upper right panel in Fig. 1).

  At  $\sin^2\theta_{23} \gtap 0.50$,
the differences between 
$N^{2\nu}_{\mu}/N^{0}_{e}$ and
$(N^{3\nu}_{\mu}/N^{3\nu}_{e})_{\rm NH,~IH}$
become noticeable 
already
at $\cos\theta_n \gtap 0.4$.
They increase with the increasing of
$\sin^2\theta_{23}$ and/or $\sin^22\theta_{13}$, and
are maximal for $\sin^2\theta_{23}= 0.64$ 
and $\sin^22\theta_{13} = 0.10$.
The dependence on $\sin^2\theta_{23}$
is particularly strong.
The deviations from the vacuum oscillation
ratio $(N^{3\nu}_{\mu}/N^{3\nu}_{e})_{vac}$
increase with the increasing of 
$\cos\theta_n \gtap 0.4$ as well;
they are maximal for the
Earth-core-crossing neutrinos,
$\cos\theta_n \gtap 0.84$.

 For $\sin^2\theta_{23}= 0.50$ 
and $\sin^22\theta_{13} = 0.05;~0.10$,
the relative difference
between $N^{2\nu}_{\mu}/N^{0}_{e}$ and
$(N^{3\nu}_{\mu}/N^{3\nu}_{e})_{\rm NH}$
in the interval
$\cos\theta_n \cong (0.5 - 0.84)$
is approximately $11\%;~13\%$;
for $\sin^22\theta_{23} = 0.64$ and
$\sin^22\theta_{13} = 0.05;~0.10$,
it is $18\%;~24\%$.
The same differences are larger in the 
Earth core interval $\cos\theta_n \cong (0.84 - 1.0)$
due to the mantle-core enhancement (NOLR) \cite{SP3198,106,107},
reaching on average values of $24\%;~27\%$ 
for $\sin^2\theta_{23}= 0.50$ 
and $\sin^22\theta_{13} = 0.05;~0.10$,
and $ 35\%;~38\%$ 
for $\sin^22\theta_{23} = 0.64$ and
the same two values of $\sin^22\theta_{13}$.

 For  $\sin^2\theta_{23} = 0.50$, 
and $\sin^22\theta_{13} = 0.05;~0.10$,
the relative difference 
between $N^{2\nu}_{\mu}/N^{0}_{e}$ and
the ratio $N_{\mu}/N_e$
in the case of IH
neutrino mass spectrum, 
$(N^{3\nu}_{\mu}/N^{3\nu}_{e})_{\rm IH}$,
in the interval $\cos\theta_n \cong (0.50 - 0.84)$
has a mean value of approximately $6\%;~8\%$, 
while if $\sin^22\theta_{23} = 0.64$
it is approximately $10\%;~14\%$.
It is larger in the Earth core bin,
$\cos\theta_n \cong (0.84 - 1.0)$,
reaching approximately 
$25\%$ for $\sin^22\theta_{23} = 0.64$
and $\sin^22\theta_{13} = 0.10$.

  The magnitude of the difference between
$N^{2\nu}_{\mu}/N^{0}_{e}$ and
$(N^{3\nu}_{\mu}/N^{3\nu}_{e})_{\rm NH(IH)}$
exhibits a relatively strong dependence on
the minimal value of the neutrino energy $E$
from the integration interval, $E_{min}$, and a rather mild
dependence on the maximal $E$ in the interval,
$E_{max}$. 
With the increase of $E_{min}$,
the relative magnitude of the contributions 
to the energy-integrated event rates of interest,
coming from the energy interval in which 
the matter effects are significant,
also increases, leading to larger
difference between
$N^{2\nu}_{\mu}/N^{0}_{e}$ and
$(N^{3\nu}_{\mu}/N^{3\nu}_{e})_{\rm NH(IH)}$.
This is illustrated in Fig. 2,
where the ratios of interest are shown
as functions of $\cos\theta_n$ 
for $|\Delta m^2_{31}| = 3\times 10^{-3}~{\rm eV^2}$,
$\sin^2\theta_{23} = 0.50$,
$\sin^2\theta_{13} = 0.10$,
and i) $E_{min} = 4$ GeV,
$E_{max} = 10$ GeV (left panel), and
ii) $E_{min} = 2$ GeV, $E_{max} = 100$ GeV (right panel). 
Increasing $E_{min}$ ($E_{max}$) while keeping
$E_{max}$ ($E_{min}$) intact would lead to
the decreasing (increasing) of the 
statistics in the samples of $\mu-$like and
$e-$like events of interest.

 As Fig. 2 illustrates, the 
relative differences between
i) $N^{2\nu}_{\mu}/N^{0}_{e}$ and
$(N^{3\nu}_{\mu}/N^{3\nu}_{e})_{\rm NH}$
and ii) $N^{2\nu}_{\mu}/N^{0}_{e}$
and $(N^{3\nu}_{\mu}/N^{3\nu}_{e})_{\rm IH}$,
increase noticeably in the interval
$\cos\theta_n \cong (0.40 - 0.65)$
with the increase of $E_{min}$,
being almost constant for $E_{min} = 4$ GeV
and reaching the values of approximately $29\%$ and $19\%$, 
respectively
\footnote{The relative difference between
$(N^{3\nu}_{\mu}/N^{3\nu}_{e})_{\rm IH}$
and $(N^{3\nu}_{\mu}/N^{3\nu}_{e})_{\rm NH}$ is
$\sim 12\%$.}.
For $\cos\theta_n \cong (0.84 - 1.0)$,
the differences under discussion exhibit 
a characteristic oscillatory pattern.
These differences have a completely different behavior 
as functions of $\cos\theta_n$ if $E_{max}$ is increased
to 100 GeV, keeping $E_{min} = 2$ GeV
(Fig. 2, right panel): they increase 
approximately linearly with $\cos\theta_n$,
starting from 0 at $\cos\theta_n \cong 0.25$,
and having at $\cos\theta_n \cong 0.70;~0.84$
the values i) $17\%;~21\%$
and ii) $9\%;~13\%$, respectively.

\begin{figure}
\includegraphics[height=7.2cm]{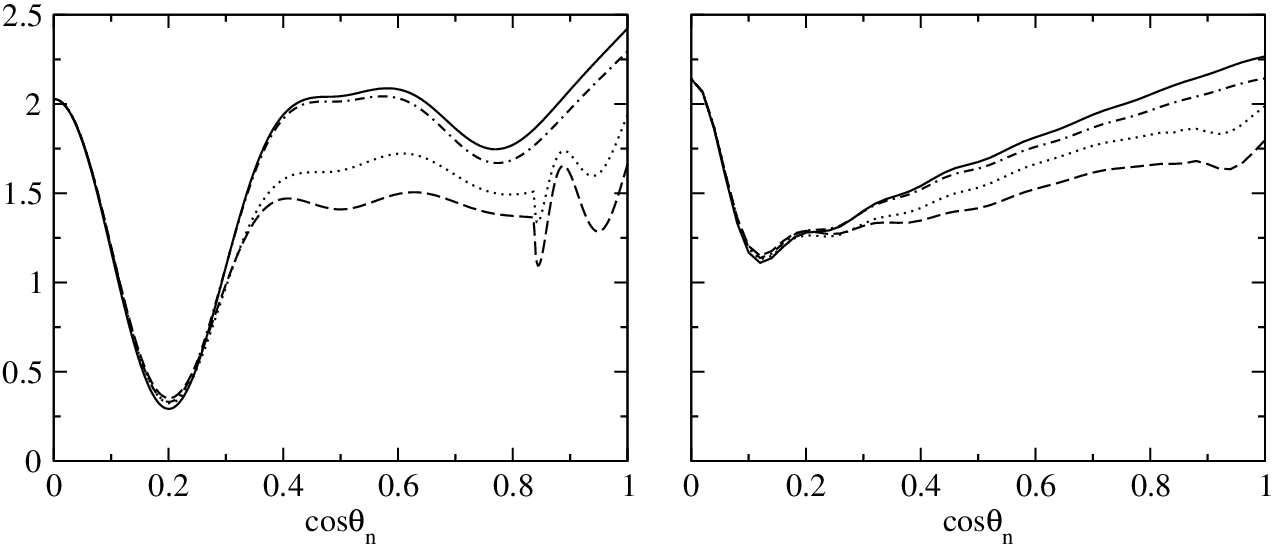}
\caption{\label{nadir} 
The same as in Fig. 1
but for the ratios of the 
$\mu-$ and $e-$ like events (or event rates), 
integrated respectively over the neutrino energy 
in the intervals $E = (4.0 - 10.0)$ GeV (left panel),
and $E = (2.0 - 100.0)$ GeV (right panel), and
for $|\Delta m^2_{31}| = 3\times 10^{-3}~{\rm eV^2}$,
$\sin^2\theta_{23} = 0.50$ and
$\sin^22\theta_{13} = 0.10$.
}   
\end{figure}

\begin{figure}
\includegraphics[height=18.2cm]{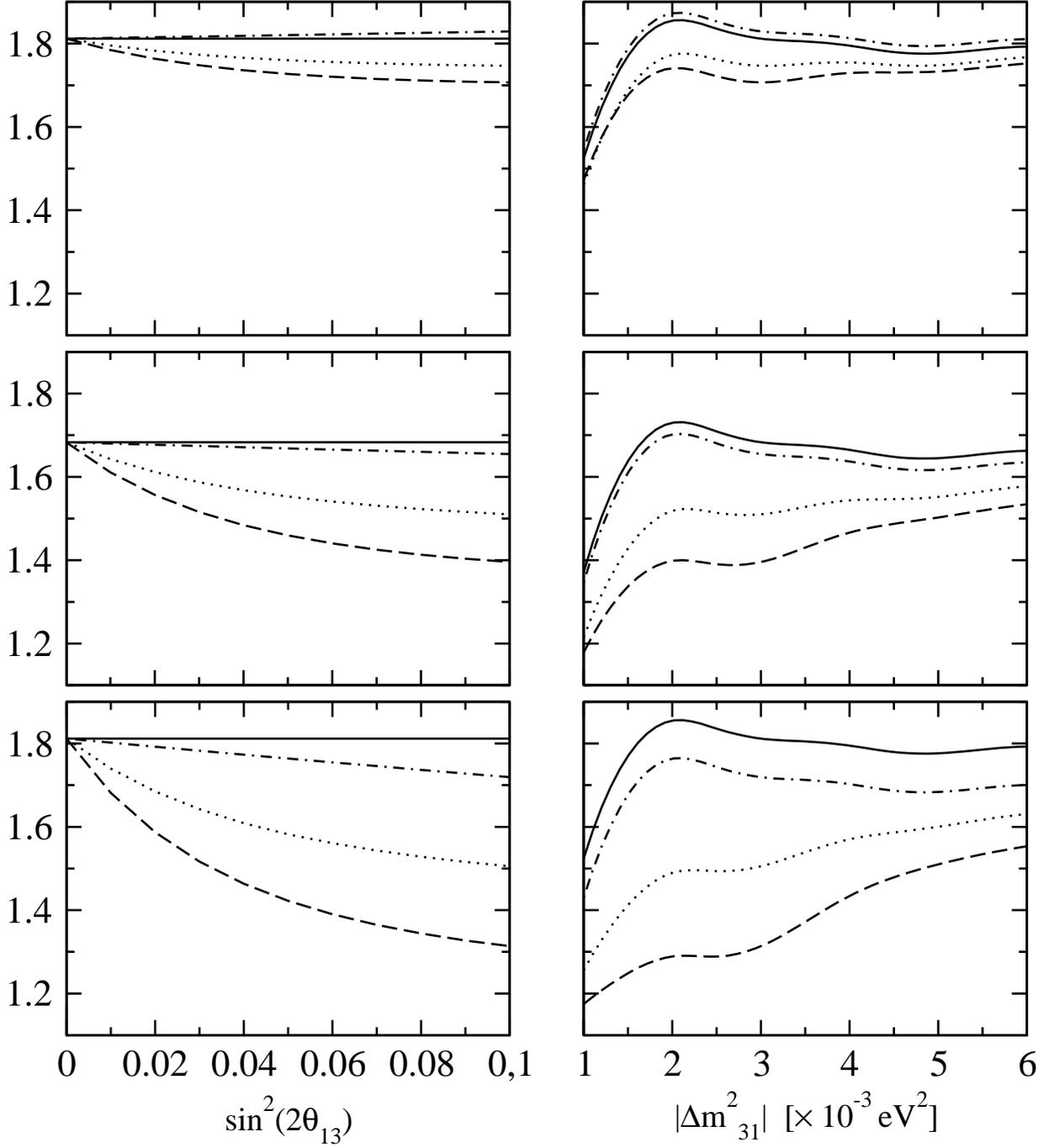}
\caption{
 The four different ratios of the multi-GeV
$\mu-$ and $e-$ like events (or event rates), 
integrated over the neutrino energy 
in the interval $E = (2.0 - 10.0)$ GeV and over
the Nadir angle in the interval corresponding to
$0.40 \leq \cos{\theta_n} \leq 1.0$,
as functions i) of $\sin^22\theta_{13}$ 
for $|\Delta m^2_{31}| = 3\times 10^{-3}~{\rm eV^2}$ 
(left panels), and ii) of 
$|\Delta m^2_{31}|$ for 
$\sin^22\theta_{13} = 0.10$ (right panels):
$N^{2\nu}_{\mu}/N^{0}_{e}$ (solid lines),
$(N^{3\nu}_{\mu}/N^{3\nu}_{e})_{vac}$ (dash-dotted lines),
$(N^{3\nu}_{\mu}/N^{3\nu}_{e})_{\rm NH}$ (dashed lines) and
$(N^{3\nu}_{\mu}/N^{3\nu}_{e})_{\rm IH}$ (dotted lines).
The results shown are obtained
for $\sin^2\theta_{23} = 0.36~(\rm upper~panels);~0.50
~(\rm middle~panels);~0.64~(\rm lower~panels)$.
}
\end{figure}

   In Fig. 3 we show the
predictions for the different 
ratios of $\mu-$like and $e-$like 
event rates we consider,
integrated over the neutrino energy
in the interval $E = (2 - 10)$ GeV and over
$\cos\theta_n$ in the interval
$\cos\theta_n = (0.4 - 1.0)$,
i) as functions of $\sin^22\theta_{13}$
for $|\Delta m^2_{31}| = 3\times 10^{-3}~{\rm eV^2}$
(left panels), and ii) as functions of 
$|\Delta m^2_{31}|$ for
$\sin^22\theta_{13} = 0.10$ (right panel).
In each case the results presented
are for three values of 
$\sin^2\theta_{23} = 0.36;~ 0.50;~0.64$.
The differences between the
ratios of the integrated 
$\mu-$like and $e-$like event rates 
of interest, i.e.,
i) between $N^{2\nu}_{\mu}/N^{0}_{e}$ and
$(N^{3\nu}_{\mu}/N^{3\nu}_{e})_{\rm NH}$
and ii) between $N^{2\nu}_{\mu}/N^{0}_{e}$
and $(N^{3\nu}_{\mu}/N^{3\nu}_{e})_{\rm IH}$,
increase rather rapidly as 
$\sin^22\theta_{13}$ increases
from 0 to $\sin^22\theta_{13} \cong 0.05$,
while the increase is slower 
in the interval $\sin^22\theta_{13} \cong (0.05 - 0.10)$.
At $\sin^22\theta_{13} \cong 0.05$,
the relative differences 
between $N^{2\nu}_{\mu}/N^{0}_{e}$ and
i) $(N^{3\nu}_{\mu}/N^{3\nu}_{e})_{\rm NH}$
and ii) $(N^{3\nu}_{\mu}/N^{3\nu}_{e})_{\rm IH}$
in the case under discussion
are respectively $13\%$ and $8\%$ 
for $\sin^2\theta_{23} = 0.50$;
for $\sin^2\theta_{23} = 0.64$
they are considerably larger,
$21\%$ and $13\%$.
If $\sin^22\theta_{13} \cong 0.10$,
the same two differences 
for $\sin^2\theta_{23} = 0.50;~0.64$
read $17\%;~28\%$ and $10\%;~17\%$, 
respectively.

  It follows from Fig. 3, right panel, that
differences between  the integrated ratios
$N^{2\nu}_{\mu}/N^{0}_{e}$
(or $(N^{3\nu}_{\mu}/N^{3\nu}_{e})_{vac}$) and
$(N^{3\nu}_{\mu}/N^{3\nu}_{e})_{\rm NH}$,
and $(N^{3\nu}_{\mu}/N^{3\nu}_{e})_{\rm IH}$
and $(N^{3\nu}_{\mu}/N^{3\nu}_{e})_{\rm NH}$,
are maximal for values of $|\Delta m^2_{31}|$
lying in the interval 
$(2 - 3)\times 10^{-3}~{\rm eV^2}$,
which are favored by the current 
atmospheric neutrino data.

\begin{figure}
\includegraphics[height=7.2cm]{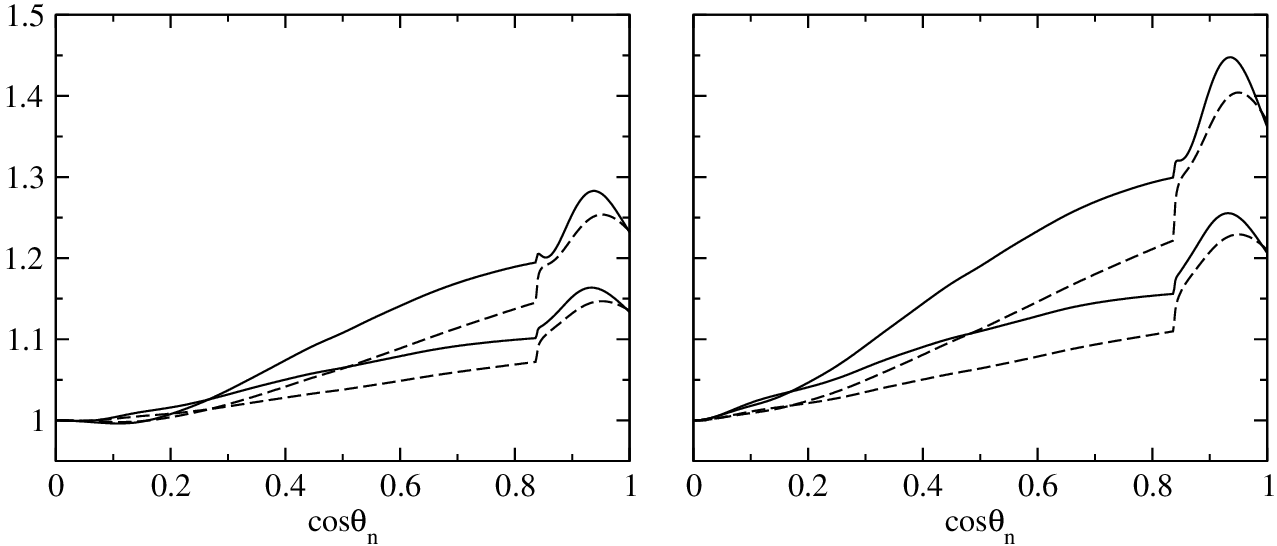}
\caption{The ratio $N^{3\nu}_e/N^{o}_e$
of $e-$like multi-GeV events (or event rates),
for $\nu_e$ and $\bar{\nu}_e$
taking part in 3-neutrino oscillations in the Earth 
($N^{3\nu}_e$), and $\nu_e$ and $\bar{\nu}_e$
not taking part in the oscillations ($N^{o}_e$),
as a function of $\cos{\theta_n}$ for
$\sin^{2}{2\theta_{13}} = 0.10~({\rm solid~lines});
~0.05$ (dashed lines), and for neutrino mass spectrum
with normal hierarchy (upper solid or dashed lines) 
and with inverted hierarchy (lower solid or dashed lines).
The results shown are for
$\sin^{2}{\theta_{23}} = 0.5~({\rm left~panel});~
0.64$ (right panel), and for 
$|\Delta m^{2}_{31}| = 3 \times 10^{-3} \mbox{eV}^2$.
See text for details.
}
\end{figure}

\begin{figure}
\includegraphics[height=12cm]{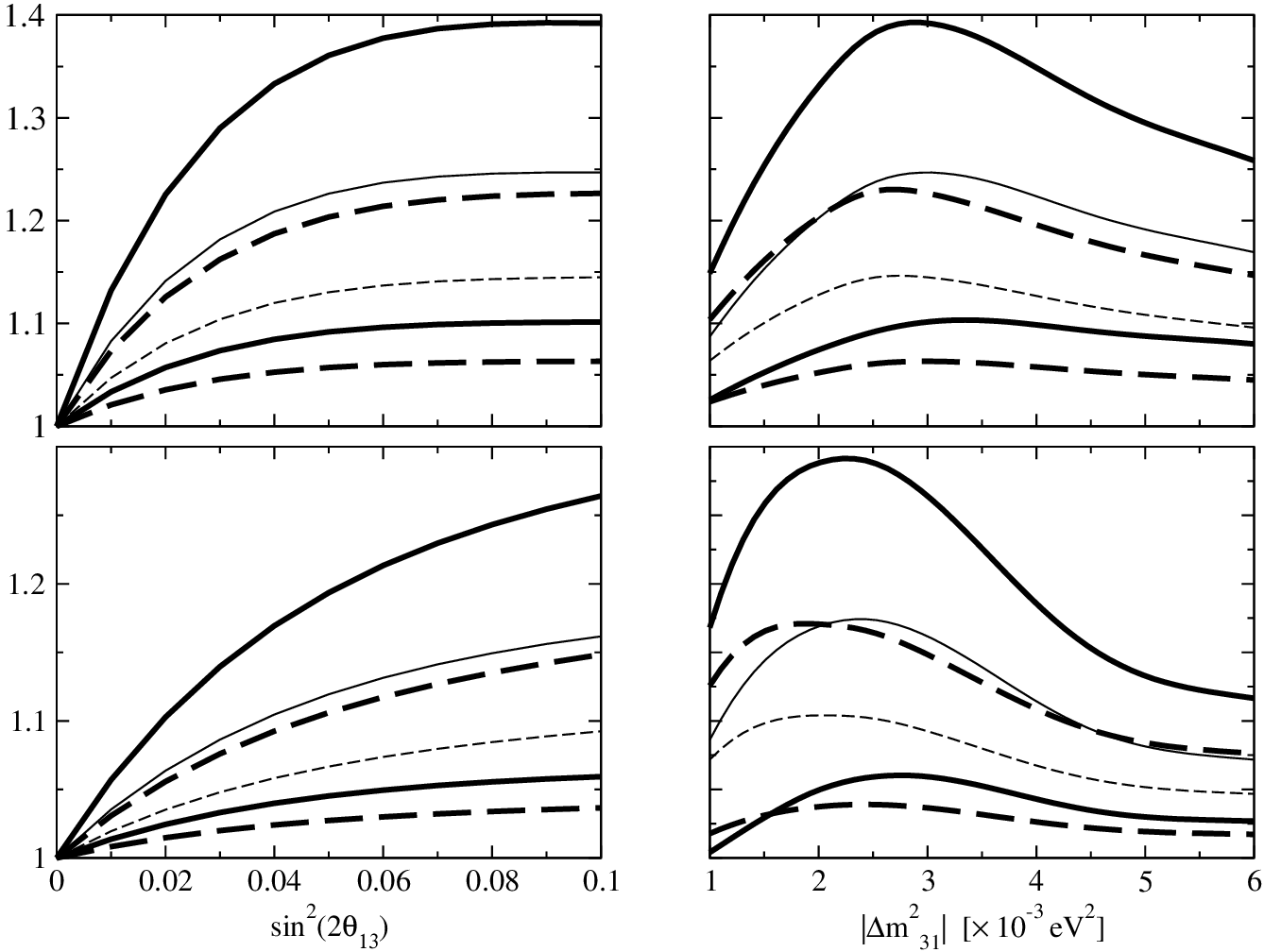}
\caption{ The dependence of the ratio
$N^{3\nu}_e/N^{o}_e$ i) on $\sin^{2}{\theta_{13}}$
  for $|\Delta m^{2}_{31}| = 3 \times 10^{-3} \mbox{eV}^2$
  (left panels), and ii) on $|\Delta m^{2}_{31}|$ for
  $\sin^{2}{\theta_{13}} = 0.10$ (right panels), 
in the cases of neutrino mass spectrum with normal hierarchy
(solid lines) and inverted hierarchy (dashed lines).
In this figure $N^{3\nu}_e$ and $N^{o}_e$ 
are the multi-GeV e-like events (or event rates)
for $\nu_e$ and $\bar{\nu}_e$
taking part in 3-neutrino oscillations in the Earth 
and $\nu_e$ and $\bar{\nu}_e$
not taking part in the oscillations, respectively,
integrated over the Nadir angle $\theta_n$
in the intervals corresponding to 
a) $0.84 \leq \cos\theta_n \leq 1.0$ - neutrinos
crossing the Earth core (upper panels), and 
b) $0.40 \leq \cos\theta_n \leq 1.0$ (lower panels).
The results shown are for 
$\sin^2\theta_{23} = 0.36$ (lower doubly thick solid or dashed lines);
0.50 (middle thin solid or dashed lines); 0.64 (upper doubly thick 
solid or dashed lines).
}
\end{figure}

     In Figs. 4 and 5 we present results just for the
multi-GeV e-like event rate. The Earth matter 
effects are largest
in the oscillations of the atmospheric $\nu_e$ or
$\bar{\nu}_e$. The dependence of the ratio
$N^{3\nu}_{e}/N^{0}_e$, 
$N^{3\nu}_{e}$ and $N^{0}_{e}$
being the numbers of multi-GeV $e$-like 
events (or event rates) 
predicted in the cases of 3-$\nu$ 
oscillations of $\nu_e$, $\bar{\nu}_e$, 
and of absence of oscillations
($\sin^2\theta_{13} = 0$),
on $\cos\theta_n$ is shown in Fig. 4
for $\sin^2\theta_{23} = 0.50;~0.64$ and
for $\sin^2\theta_{13} = 0.05;~0.10$.
The deviation of $N^{3\nu}_{e}/N^{0}_e$
from 1 would signal that 
$\nu_e$ and $\bar{\nu}_e$ take part in 
oscillations and that 
$\sin^2\theta_{13} \neq 0$.
We can have $N^{3\nu}_{e}/N^{0}_e > 1$
only if a substantial fraction of the
atmospheric $\nu_{\mu}$ (or 
$\bar{\nu}_{\mu}$) oscillate
into $\nu_{e}$ ($\bar{\nu}_{e}$).
As Fig. 4 shows, the ratio
$N^{3\nu}_{e}/N^{0}_e$
increases with the increasing of
$\cos\theta_n$ and can be 
significantly greater than 1 
for $\cos\theta_n \gtap 0.4$.
At $\cos\theta_n = 0.8$, for instance,
we have for $\sin^2\theta_{23} = 0.50$ and
$\sin^22\theta_{13} = 0.05;~0.10$
in the case of NH neutrino mass spectrum
$(N^{3\nu}_{e}/N^{0}_e)_{NH} \cong 1.14;~1.19$,
while for $\sin^2\theta_{23} = 0.64$ one
finds $(N^{3\nu}_{e}/N^{0}_e)_{NH} \cong 1.21;~1.29$.
For IH neutrino mass spectrum the  
ratio of interest, $(N^{3\nu}_{e}/N^{0}_e)_{IH}$,
is smaller and the corresponding values
read: 
$(N^{3\nu}_{e}/N^{0}_e)_{IH} \cong 1.07;~1.10$ and
$(N^{3\nu}_{e}/N^{0}_e)_{IH} \cong 1.11;~1.15$, respectively.
For the Earth-core-crossing neutrinos both
$(N^{3\nu}_{e}/N^{0}_e)_{NH}$ and
$(N^{3\nu}_{e}/N^{0}_e)_{IH}$ 
are larger due to the NOLR effect,
and for $\sin^2\theta_{23} = 0.64$
and $\sin^2\theta_{13} = 0.10$
reach the values 
$(N^{3\nu}_{e}/N^{0}_e)_{NH} \cong 1.45$ and
$(N^{3\nu}_{e}/N^{0}_e)_{IH} \cong 1.26$.

  For given $\sin^2\theta_{23}$ and
and $\sin^2\theta_{13}$,
the maximum of the ratio of the multi-GeV e-like event
rates of interest $N^{3\nu}_{e}$ and $N^{0}_e$, 
integrated over $\theta_n$ in the 
intervals corresponding to
a) $\cos\theta_n = (0.84 - 1.0)$ (Earth core bin)
and b) $\cos\theta_n = (0.40 - 1.0)$, 
occurs again for values of 
$|\Delta m^2_{31}| = (2 - 3)\times 10^{-3}~{\rm eV^2}$
(Fig. 5, right panels),
favored by the existing atmospheric neutrino data.
For $|\Delta m^2_{31}| = 3\times 10^{-3}~{\rm eV^2}$, 
$\sin^2\theta_{13} = (0.05 - 0.10)$
and $\sin^2\theta_{23} = 0.64$, 
the two types of $\theta_n-$ 
integrated ratios can be as large as: 
case a) $(N^{3\nu}_{e}/N^{0}_e)_{NH(IH)} 
\cong 1.36 - 1.39~(1.20 - 1.23)$, 
and case b)
$(N^{3\nu}_{e}/N^{0}_e)_{NH(IH)} \cong 1.19 - 1.26~(1.11 - 1.15)$
(Fig. 5, left and right panels).

\section{Conclusions}

\hskip 0.6truecm  In the present article we 
have studied the possibility to obtain
evidences for Earth matter enhanced 
3-neutrino oscillations of the
atmospheric neutrinos involving, in particular,
the $\nu_e$ (or $\bar{\nu}_e$),
from the analysis of the 
$\mu-$like and $e-$like multi-GeV
event data accumulated by the SK experiment,
or that can be provided by 
future water-\v{C}erenkov detectors.
Such evidences would give also
important quantitative information on the values 
of $\sin^2\theta_{13}$ and 
$\sin^2\theta_{23}$ and on the sign of
$\deltaatm = \Delta m^2_{31}$.
We have considered 3-neutrino oscillations 
of the atmospheric $\nu_{\mu}$, $\bar{\nu}_{\mu}$,
$\nu_e$ and $\bar{\nu}_e$,
assuming that the inequality 
$\deltasol = \Delta m^2_{21} \ll 
|\deltaatm| = |\Delta m^2_{31}|$
holds, as is suggested 
by the current solar and atmospheric 
neutrino data.
Depending on the sign of 
$\Delta m^2_{31}$,
the Earth matter effects 
in this case can enhance either
the  $\nu_{\mu} \rightarrow \nu_e$ and
$\nu_{e} \rightarrow \nu_{\mu}$,
or the $\bar{\nu}_{\mu} \rightarrow \bar{\nu}_e$
and $\bar{\nu}_{\mu} \rightarrow \bar{\nu}_{\mu}$
transitions if $\sin^2\theta_{13}\neq 0$.
The effects of the enhancement 
can be substantial for $\sin^2\theta_{13}\gtap 0.01$
and $\sin^2\theta_{23}\gtap 0.50$.
They are largest in
the multi-GeV $e$-like and $\mu-$like 
samples of events and for atmospheric 
neutrinos with relatively large path length
in the Earth, crossing deeply the mantle
or the mantle and the core, i.e., for
$\cos\theta_n \gtap 0.4$, where $\theta_n$ 
is the Nadir angle characterizing 
the neutrino trajectory in the Earth.

 As observables which are 
particularly sensitive 
to the Earth matter effects, and 
thus to the values
of $\sin^2\theta_{13}$ and
$\sin^2\theta_{23}$,
and to the sign of 
$\Delta m^2_{31}$,
we have considered the
Nadir-angle distributions of the 
ratios $N^{3\nu}_{\mu}/N^{3\nu}_{e}$ and
$N^{3\nu}_{e}/N^{0}_e$,
where $N^{3\nu}_{\mu}$ and $N^{3\nu}_{e}$       
are the multi-GeV $\mu-$like and  $e$-like 
numbers of events (or event rates) 
in the case of 3-$\nu$ 
oscillations of the atmospheric $\nu_e$, $\bar{\nu}_e$
and $\nu_{\mu}$, $\bar{\nu}_{\mu}$, and
$N^{0}_{e}$ is the number of $e-$like events 
in the case of absence of oscillations
($\sin^2\theta_{13} = 0$).
As is well-known, 
the ratio of the energy and Nadir angle integrated
$\mu-$like and $e-$like events, $N_{\mu}/N_e$, 
has been measured with a 
relatively high precision by the
SK experiment \cite{SKatm00}.

  We have obtained predictions for the 
Nadir-angle distributions of
$N^{3\nu}_{\mu}/N^{3\nu}_{e}$ and of 
$N^{3\nu}_{e}/N^{0}_e$ 
both for neutrino mass spectra with 
normal ($\Delta m^2_{31} > 0$)
and inverted  ($\Delta m^2_{31} < 0$) hierarchy,
$(N^{3\nu}_{\mu}/N^{3\nu}_{e})_{\rm NH}$, 
$(N^{3\nu}_{\mu}/N^{3\nu}_{e})_{\rm IH}$,
$(N^{3\nu}_{e}/N^{0}_{e})_{\rm NH}$ and
$(N^{3\nu}_{e}/N^{0}_{e})_{\rm IH}$, 
and for $\sin^2\theta_{23} = 0.64;~0.50;~0.36$.
We compared the latter with the predicted
Nadir-angle distributions i) of the ratio
$N_{\mu}/N_{e}$
for the case the 3-neutrino
oscillations taking place in vacuum,
$(N^{3\nu}_{\mu}/N^{3\nu}_{e})_{\rm vac}$, and
ii) of the ratio $N^{2\nu}_{\mu}/N^{0}_{e}$,
the predicted number of $\mu-$like
and $e-$like events in the case of 2-neutrino
$\nu_{\mu} \rightarrow \nu_{\tau}$ and
$\bar{\nu}_{\mu} \rightarrow \bar{\nu}_{\tau}$
oscillations of the atmospheric 
$\nu_{\mu}$ and $\bar{\nu}_{\mu}$,
and $\nu_{e}$ and $\bar{\nu}_{e}$ 
not taking part in the oscillations
($\sin^2\theta_{13} = 0$).
The dependence of the 
Nadir angle distributions
of $(N^{3\nu}_{\mu}/N^{3\nu}_{e})_{\rm NH}$, 
$(N^{3\nu}_{\mu}/N^{3\nu}_{e})_{\rm IH}$,
$(N^{3\nu}_{\mu}/N^{3\nu}_{e})_{\rm vac}$ and
$N^{2\nu}_{\mu}/N^{0}_{e}$
on the minimal and maximal neutrino energies 
in the energy integration interval  
has also been studied.
Predictions for these 
four different types of 
ratios of the suitably integrated
over $\theta_n$
Nadir angle distributions
of the $\mu-$like and $e-$like 
multi-GeV events were also derived.
 
   Our results are presented graphically in Figs. 1 - 5.
We find that for $\sin^2\theta_{23}= 0.50$ 
and $\sin^22\theta_{13} = 0.05$,
the relative difference
between $N^{2\nu}_{\mu}/N^{0}_{e}$ and
$(N^{3\nu}_{\mu}/N^{3\nu}_{e})_{\rm NH}$
in the interval
$\cos\theta_n \cong (0.5 - 0.84)$
is approximately $11\%$. If
$\sin^22\theta_{23} = 0.64$ and
$\sin^22\theta_{13} = 0.10$, the same 
difference is approximately $24\%$ (Fig. 1).
The relative difference 
between $N^{2\nu}_{\mu}/N^{0}_{e}$ and
$(N^{3\nu}_{\mu}/N^{3\nu}_{e})_{\rm NH}$
is larger in the 
Earth core interval $\cos\theta_n \cong (0.84 - 1.0)$
due to the mantle-core enhancement (NOLR) \cite{SP3198,106,107},
reaching on average values of $24\%$ 
for $\sin^2\theta_{23}= 0.50$ 
and $\sin^22\theta_{13} = 0.05$,
and of $~38\%$ 
for $\sin^22\theta_{23} = 0.64$ and 
$\sin^22\theta_{13} = 0.10$ (Fig. 1).

 In the case of neutrino mass spectrum
with inverted hierarchy, 
the relative difference 
between $N^{2\nu}_{\mu}/N^{0}_{e}$ and
$(N^{3\nu}_{\mu}/N^{3\nu}_{e})_{\rm IH}$
in the interval $\cos\theta_n \cong (0.50 - 0.84)$
has a mean value of approximately $14\%$,
for  $\sin^2\theta_{23} = 0.64$
and $\sin^22\theta_{13} = 0.10$.
It reaches approximately 
$25\%$ in the Earth core bin,
$\cos\theta_n \cong (0.84 - 1.0)$,
for these values of 
$\sin^22\theta_{23}$
and $\sin^22\theta_{13}$ (Fig. 1).

  The magnitude of the difference between
$N^{2\nu}_{\mu}/N^{0}_{e}$ and
$(N^{3\nu}_{\mu}/N^{3\nu}_{e})_{\rm NH(IH)}$
exhibits a relatively strong dependence on
the minimal value of the neutrino energy $E$
from the integration interval, $E_{min}$, 
and a rather mild dependence on the maximal $E$ 
in the integration interval,
$E_{max}$ (Fig. 2). 
Increasing 
$E_{min}$ from 2 GeV to 4 GeV and keeping
$E_{max} = 10$ GeV, leads for 
$|\Delta m^2_{31}| = 3\times 10^{-3}~{\rm eV^2}$,
$\sin^2\theta_{23} = 0.50$ and
$\sin^2\theta_{13} = 0.10$
to a considerably larger 
relative difference between
$N^{2\nu}_{\mu}/N^{0}_{e}$ and
$(N^{3\nu}_{\mu}/N^{3\nu}_{e})_{\rm NH(IH)}$
in the interval $\cos\theta_n \cong (0.40 - 0.65)$,
which reaches approximately 
$29\%~(19\%)$ (Fig. 2). This difference is larger 
also in the Earth core bin,
$\cos\theta_n \cong (0.84 - 1.0)$.
We have also found that the
differences between the Nadir angle and
energy integrated ratios
$N^{2\nu}_{\mu}/N^{0}_{e}$
(or $(N^{3\nu}_{\mu}/N^{3\nu}_{e})_{vac}$) and
$(N^{3\nu}_{\mu}/N^{3\nu}_{e})_{\rm NH}$,
and $(N^{3\nu}_{\mu}/N^{3\nu}_{e})_{\rm IH}$
and $(N^{3\nu}_{\mu}/N^{3\nu}_{e})_{\rm NH}$,
are maximal for values of $|\Delta m^2_{31}|$
lying in the interval 
$(2 - 3)\times 10^{-3}~{\rm eV^2}$,
which are favored by the current 
atmospheric neutrino data (Fig. 3).
The same conclusion is valid for 
the integrated ratio of
$e-$like events
$N^{3\nu}_{e}/N^{0}_e$ (Fig. 5).    
 
  We have also shown that 
in the case of the 3-neutrino oscillations 
of the atmospheric neutrinos considered,
the ratio $N^{3\nu}_{e}/N^{0}_e$
increases with the increasing of
$\cos\theta_n$ and can be 
significantly greater than 1 
for $\cos\theta_n \gtap 0.4$.
At $\cos\theta_n = 0.8$, for instance,
we have for $\sin^2\theta_{23} = 0.50$ and
$\sin^22\theta_{13} = 0.10$
in the case of NH neutrino mass spectrum
$(N^{3\nu}_{e}/N^{0}_e)_{NH} \cong 1.19$,
while for $\sin^2\theta_{23} = 0.64$ one
finds $(N^{3\nu}_{e}/N^{0}_e)_{NH} \cong 1.29$ (Fig. 4).
For IH neutrino mass spectrum the  
ratio of interest, $(N^{3\nu}_{e}/N^{0}_e)_{IH}$,
is smaller: 
$(N^{3\nu}_{e}/N^{0}_e)_{IH} \cong 1.10~{\rm and}~1.15$.
For the Earth-core-crossing neutrinos both
$(N^{3\nu}_{e}/N^{0}_e)_{NH}$ and
$(N^{3\nu}_{e}/N^{0}_e)_{IH}$ 
are enhanced due to the NOLR effect,
and for $\sin^2\theta_{23} = 0.64$
and $\sin^2\theta_{13} = 0.10$
reach the values $(N^{3\nu}_{e}/N^{0}_e)_{NH} 
\cong 1.45$ and
$(N^{3\nu}_{e}/N^{0}_e)_{IH} \cong 1.26$
(Fig. 4).

 It follows from our results that the 
Earth matter effects in the Nadir angle 
distribution of the ratio
of the multi-GeV $\mu-$like and $e-$like atmospheric 
neutrino events (or event rates),
measured in the Super-Kamiokande
(or any future water-\v{C}erenkov) experiment, 
might be observable if 
the atmospheric neutrinos, including the
$\nu_e$ and $\bar{\nu}_e$,
take part in 3-neutrino oscillations
and $\sin^2\theta_{13}$ and
$\sin^2\theta_{23}$ are sufficiently
large. The observation of 
relatively large Earth matter effects 
(20\% - 30\%) at $\cos\theta_n \gtap 0.4$
would clearly indicate that
$\sin^2\theta_{13} \gtap 0.01$,
$\sin^2\theta_{23} \gtap 0.50$,
and would suggest
that the neutrino mass spectrum
is with normal hierarchy,
$\Delta m^2_{31} > 0.$
However, distinguishing statistically 
between the neutrino mass 
spectrum with normal and inverted hierarchy
requires a high precision measurement 
of the Nadir angle distribution
of the multi-GeV ratio
$N_{\mu}/N_e$ and is a rather
challenging task.
      
\section*{Acknowledgements}
This work is supported in part 
by the Italian MIUR and INFN under the programs 
``Fenome-nologia delle Interazioni Fondamentali'' 
and ``Fisica Astroparticellare'',
by the U.S. National Science
Foundation under Grant No. PHY99-07949 (S.T.P.),
by the Spanish Grant FPA2002-00612 of the MCT 
(J.B. and S.P.-R.) and 
by the Spanish MECD for a
FPU fellowship (S.P.-R.).
S.T.P. would like to thank the organizers of the Program
on ``Neutrinos: Data, Cosmos and the Planck Scale''
at KITP, Univ. of California at Santa Barbara, 
where part of the  work on the present study was done,
for kind hospitality, and M. Freund 
for discussions at the initial stage of this work.
S.P.-R. would like to thank the Elementary Particle 
Physics Sector of SISSA, Trieste, Italy,
for kind hospitality 
during part of this study.


\begin{thebibliography}{100}

\bibitem{Cl98}   B.~T.~Cleveland {\em et al.}, 
                Astrophys.\ J.\ {\bf 496} (1998) 505;
                Y.~Fukuda {\em et al.},
                Phys.\ Rev.\ Lett.\  {\bf 77} (1996) 1683;
                V.~Gavrin, Nucl. Phys. Proc. Suppl. {\bf 91} (2001) 36;
                W.~Hampel {\em et al.},
                Phys.\ Lett.\ B {\bf 447} (1999) 127;
                M.~Altmann {\em et al.},
                Phys.\ Lett.\ B {\bf 490} (2000) 16.

\bibitem{SKsol} Super-Kamiokande Collaboration,
                Y.~Fukuda {\em et al.}, 
                Phys.\ Rev.\ Lett.\  {\bf 86} (2001) 5651 and 5656.

\bibitem{SKatm00} Super-Kamiokande Collaboration,
                 M.~Shiozawa,
                 talk given at the Int. Conf. on Neutrino Physics and 
                 Astrophysics ``Neutrino'02'', May 25 - 30, 2002, 
                 Munich, Germany.

\bibitem{SNO1} SNO Collaboration,
               Q.~R.~Ahmad {\it et al.}, 
               Phys.\ Rev.\ Lett.\ {\bf 87} (2001) 071301.


\bibitem{SNO2} SNO Collaboration,
               Q.~R.~Ahmad \textit{et al.}, 
               Phys.\ Rev.\ Lett.\  {\bf 89} (2002) 011302 and 011301. 

\bibitem{KamLAND}
KamLAND Collaboration, K.~Eguchi {\it et al.},
Phys.\ Rev.\ Lett.\  {\bf 90} (2003) 021802.

\bibitem{Pont4667}
B.~Pontecorvo, Chalk River Lab. report PD--205, 1946;
Zh. Eksp. Teor. Fiz. {\bf 53} (1967) 1717.

\bibitem{Davis68}
R.~Davis, D.~S.~Harmer and K.~C.~Hoffman,
Phys. Rev. Lett. {\bf 20}, 1205 (1968); Acta Physica Acad.
Sci. Hung. {\bf 29} Suppl. 4, 371 (1970); R. Davis,
Proc. of the ``Neutrino~`72''~Int. Conference, Balatonfured,
Hungary, June 1972 (eds. A.~Frenkel and G.~Marx,  OMKDK-TECHNOINFORM,
Budapest, 1972), p. 5.

\bibitem{BGG99}
S.~M.~Bilenky, C.~Giunti and W.~Grimus,
Prog.\ Part.\ Nucl.\ Phys.\ {\bf 43} (1999) 1.

\bibitem{BPont57} B.~Pontecorvo, 
                  Zh.\ Eksp.\ Teor.\ Fiz.\ {\bf 33} (1957) 549 
                and {\bf 34} (1958) 247;
Z.~Maki, M.~Nakagawa and S.~Sakata, 
Prog.\ Theor.\ Phys.\  {\bf 28} (1962) 870.

\bibitem{BHP80} S.~M.~Bilenky {\it et al.},
              Phys.\ Lett.\  B{\bf 94} (1980) 495;
M.~Doi {\it et al.},
Phys.\ Lett.\  \textbf{B102} (1981) 323;
J.~Bernab\'{e}u and P.~Pascual,
Nucl.\ Phys.\ B {\bf 228} (1983) 21;
P.~Langacker {\it et al.},
Nucl.\ Phys.\ B {\bf 282} (1987) 589.

\bibitem{BiPet87} S.~M.~Bilenky and S.~T.~Petcov,
                Rev.\ Mod.\ Phys.\  {\bf 59} (1987) 671. 

\bibitem{CHOOZ} M.~Apollonio {\it et al.}, 
                Phys.\ Lett.\ B {\bf 466} (1999) 415.

\bibitem{PaloV} 
F.~Boehm {\it et al.}, 
Phys.\ Rev.\ Lett.\   {\bf 84} (2000) 3764 and
Phys.\ Rev.\ D {\bf 62} (2000) 072002.

\bibitem{fogli}
G.~L.~Fogli {\it et al.},
Phys.\ Rev.\ D {\bf 67} (2003) 073002.


\bibitem{others} 
V.~Barger and D.~Marfatia,
Phys.\ Lett.\ B {\bf 555} (2003) 144;
M.~Maltoni, T.~Schwetz, and J.~W.~F.~Valle, hep-ph/0212129;
A.~Bandyopadhyay {\it et al.},
Phys.\ Lett.\ B {\bf 559} (2003) 121;
J.~N.~Bahcall, M.~C.~Gonz\'{a}lez-Garc\'{\i}a and C.~Pe\~{n}a-Garay,
JHEP {\bf 0302} (2003) 009;
H.~Nunokawa, W.~J.~C.~Teves, and R.~Zukanovich Funchal,
hep-ph/0212202;   
P.~Aliani {\it et al.}, hep-ph/0212212; 
P.~C.~de Holanda and A.~Y.~Smirnov,
JCAP {\bf 0302} (2003) 001;
A.~B.~Balantekin and H.~Yuksel,
J.\ Phys.\ G {\bf 29} (2003) 665.


\bibitem{fogliold} 
G.~L.~Fogli {\it el al.},
Phys.\ Rev.\ D {\bf 66} (2002) 093008


\bibitem{BNPChooz} S.~M.~Bilenky, D.~Nicolo and S.~T.~Petcov, 
Phys.\ Lett.\ B {\bf 538} (2002) 77.

\bibitem{ADE80} A.~De R\'{u}jula {\it et al.}, 
Nucl.\ Phys.\ B {\bf 168} (1980) 54.

\bibitem{SP3198}  S.~T.~Petcov, 
Phys.\ Lett.\ B {\bf 434} (1998) 321,
(E) {\it ibid.} B {\bf 444} (1998) 584.

\bibitem{102} S.~T.~Petcov, 
Nucl.\ Phys.\  B (Proc. Suppl.) {\bf 77} (1999) 93, 
hep-ph/9809587, hep-ph/9811205 and hep-ph/9907216.

\bibitem{104} M.~V.~Chizhov, M.~Maris 
and S.~T.~Petcov, hep-ph/9810501.

\bibitem{LBL} A.~De R\'{u}jula, M.~B.~Gavela and P.~Hern\'{a}ndez,
Nucl.\ Phys.\ B {\bf 547}, 21 (1999);
V.~Barger {\it et al.},
Phys.\ Rev.\ D {\bf 62}, 013004 (2000).

\bibitem{AMMS99} M.~Freund {\it et al.}, 
Nucl.\ Phys.\ B {\bf 578}  (2000) 27.


\bibitem{mantle}
M.~C.~Ba\~{n}uls, G.~Barenboim and J.~Bernab\'{e}u,
Phys.\ Lett.\ B {\bf 513}, 391 (2001).

\bibitem{mantleproc12}
J.~Bernab\'{e}u and S.~Palomares-Ruiz,
hep-ph/0112002, and
Nucl.\ Phys.\ Proc.\ Suppl.\  {\bf 110}, 339 (2002), 
hep-ph/0201090.

\bibitem{3nuKP88} P.~I.~Krastev and S.~T.~Petcov, 
Phys.\ Lett.\ B {\bf 205} (1988) 84.

\bibitem{CPother}
J.~Arafune and J.~Sato, Phys.\ Rev.\ D {\bf 55} (1997) 1653;
J.~Bernab\'{e}u, Proc.\ WIN'99, World Scientific (2000), p. 227, 
hep-ph/9904474.

\bibitem{Dick99} K.~Dick {\it et al.},      
Nucl.\ Phys.\ B {\bf 562} (1999) 29.

\bibitem{CPT}
J.~Bernab\'{e}u and M.~C.~Ba\~{n}uls,
Nucl.\ Phys.\ Proc.\ Suppl.\  {\bf 87} (2000) 315, 
hep-ph/0003299.

\bibitem{BPP1}  S.~M.~Bilenky, S.~Pascoli  and S.~T.~Petcov,
Phys.\ Rev.\ D {\bf 64} (2001) 053010; 
S.~Pascoli and  S.~T.~Petcov, 
Phys.\ Lett.\ B {\bf 544} (2002) 239; 
S.~Pascoli, S.~T.~Petcov and W.~Rodejohann,
Phys.\ Lett.\ B {\bf 558} (2003) 141.

\bibitem{LW78}	L.~Wolfenstein, 
		Phys.\ Rev.\ D {\bf 17}, 2369 (1978), and
               {\it ibid.} D {\bf 20}, 2634 (1979).


\bibitem{BPPW80}  V.~Barger {\it et al.},
		Phys.\ Rev.\ D {\bf 22}, 2718 (1980).


\bibitem{MS85}	S.~P.~Mikheyev and A.~Yu.~Smirnov, 
		Yad.\ Fiz.\ {\bf 42}, 1441 (1985) 
		[Sov.\ J.\ Nucl.\ Phys.\ {\bf 42}, 913 (1985)].


\bibitem{BGGKP99} S.~M.~Bilenky {\it et al.}, Phys.\ Lett.\ B
{\bf 465} (1999) 193.



\bibitem{MINOS} D.~Michael (MINOS Collaboration),
Talk at the Int. Conf. on Neutrino Physics and 
Astrophysics ``Neutrino'02'', May 25 - 30, 2002, Munich, Germany.
                
\bibitem{MSpironu02} M.~Spiro, Summary talk at the
Int. Conf. on Neutrino Physics and Astrophysics ``Neutrino'02'',
May 25 - 30, 2002, Munich, Germany.


\bibitem{HLM} 
V.~Barger, D.~Marfatia and K.~Whisnant,
Phys.\ Lett.\ B {\bf 560} (2003) 75;
P.~Huber, M.~Lindner and W.~Winter,
Nucl.\ Phys.\ B {\bf 654} (2003) 3


\bibitem{SPMPiai01} S.~T.~Petcov and M.~Piai, 
Phys.\ Lett.\ B {\bf 533} (2002) 94.


\bibitem{106} M.~V.~Chizhov and S.T.~Petcov, 
Phys.\ Rev.\ Lett.\ {\bf 83} (1999) 1096.



\bibitem{107} M.~V.~Chizhov and S.~T.~Petcov,
Phys.\ Rev.\ D {\bf 63} (2001) 073003. 



\bibitem{core}
J.~Bernab\'{e}u, S.~Palomares-Ruiz, A.~P\'{e}rez and S.~T.~Petcov,
Phys.\ Lett.\ B {\bf 531}, 90 (2002)



\bibitem{ConchaMal02} 
M.~C.~Gonz\'{a}lez-Garc\'{\i}a and M.~Maltoni,
Eur.\ Phys.\ J.\ C {\bf 26} (2003) 417



\bibitem{3nuSP88} S.~T.~Petcov, Phys.\ Lett.\ B {\bf 214}, 259 (1988).

\bibitem{Honda} M.~Honda {\it et al.}, Phys.\ Rev.\ D {\bf 52}, 4985
(1995).  

\bibitem{Bartol} V.~Agraval {\it et al.}, Phys.\ Rev.\ D {\bf 53},
1314 (1996).  

\bibitem{Naumov} G.~Fiorentini, V.~A.~Naumov and F.~L.~Villante,
                 Phys.\ Lett.\ B {\bf 578} (2000) 27.

\bibitem{Stacey:1977}
F.~D.~Stacey,
 {\it Physics of the Earth, 2$^{nd}$ edition},
John Wiley and Sons, London, New York, 1977.

\bibitem{PREM81} A.~D.~Dziewonski and D.~L.~Anderson, 
               Physics of the Earth and Planetary Interiors 
               {\bf 25}, 297 (1981).

\bibitem{Art2} M.~Maris and S.~T.~Petcov, 
Phys.\ Rev.\ D {\bf 56} (1997) 7444.

\bibitem{MP98:2layers} M.~Maris, Q.~Y.~Liu and S.~T.~Petcov, 
study performed in November - December of 1996 (unpublished). 

\bibitem{EarthCORE} R.~Jeanloz,
Annu. Rev. Earth Planet. Sci. {\bf{18}}, 356 (1990);
C.~J.~All{\`e}gre et al., Earth
Planet. Sci. Lett. {\bf 134}, 515 (1995);
W.~F.~McDonough and S.-s. Sun, Chemical Geology {\bf 120}, 223 (1995).

\bibitem{GRV94}
M.~Gluck, E.~Reya and A.~Vogt,
Z.\ Phys.\ C {\bf 67} (1995) 433.


\end{thebibliography}
\end{document}